\documentclass[a4paper,12pt]{article}
\pdfoutput=1
\usepackage{graphicx, rotating}
\usepackage{hyperref}
\usepackage{slashed}
\usepackage{ifpdf}
\ifx\pdfoutput\undefined
\usepackage[dvips,bookmarks=false]{hyperref}	% This is for arXiv.org
\else
\usepackage{hyperref}	% This is for pdftex
\fi
\hypersetup{colorlinks,bookmarksopen,bookmarksnumbered,citecolor=verdes,
linkcolor=blus,pdfstartview=FitH,urlcolor=rossos}
\def\hhref#1{\href{http://arxiv.org/abs/#1}{#1}} % in bibliography
\usepackage{amsmath}

\newcommand{\sfrac}[2]{#1/#2}

\newcommand{\beq}{\begin{equation}}
\newcommand{\eeq}{\end{equation}}
\newcommand{\fig}[1]{~\ref{fig:#1}}

\newcount\Mac  \Mac=1  % devo mettere Mac=1 se sto lavorando sul file Mac
\newcommand{\ifMac}[2]{\ifnum\Mac=1 #1 \else #2 \fi}
\def\putps(#1,#2)(#3,#4)#5#6{\ifnum\Mac=1 \put(#1,#2){\special{picture #5}}
\else  \put(#3,#4){\includegraphics{#6}} \fi}

\newcommand{\One}{\hbox{1\kern-.24em I}}

\newcommand{\pb}{\,{\rm pb}}
\newcommand{\GeV}{\,{\rm GeV}}
\newcommand{\TeV}{\,{\rm TeV}}
\newcommand{\MeV}{\,{\rm MeV}}

\newcommand{\eq}[1]{~{\rm(\ref{eq:#1})}}

\newcommand{\lascia}[1]{}
\makeatletter

\def\art{\@ifnextchar[{\eart}{\oart}}
\def\eart[#1]#2#3#4#5#6{{\rm #2}, {#3 #4} {\rm (#6) #5} [arXiv:{\hhref{#1}}]}
\def\hepart[#1]#2{{\rm #2, arXiv:\hhref{#1}}}
\newcommand{\oart}[5]{{\rm #1}, {#2 #3} {\rm (#5) #4}}

\newcounter{alphaequation}[equation]
\def\thealphaequation{\theequation\hbox to
0.6em{\hfil\alph{alphaequation}\hfil}}
\def\eqnsystem#1{
\def\@eqnnum{{\rm (\thealphaequation)}}
\def\@@eqncr{\let\@tempa\relax \ifcase\@eqcnt \def\@tempa{& & &} \or
  \def\@tempa{& &}\or \def\@tempa{&}\fi\@tempa
  \if@eqnsw\@eqnnum\refstepcounter{alphaequation}\fi
\global\@eqnswtrue\global\@eqcnt=0\cr}
\refstepcounter{equation} \let\@currentlabel\theequation \def\@tempb{#1}
\ifx\@tempb\empty\else\label{#1}\fi
\refstepcounter{alphaequation}
\let\@currentlabel\thealphaequation
\global\@eqnswtrue\global\@eqcnt=0 \tabskip\@centering\let\\=\@eqncr
$$\halign to \displaywidth\bgroup \@eqnsel\hskip\@centering
$\displaystyle\tabskip\z@{##}$&\global\@eqcnt\@ne
\hskip2\arraycolsep\hfil${##}$\hfil& \global\@eqcnt\tw@\hskip2\arraycolsep
$\displaystyle\tabskip\z@{##}$\hfil
\tabskip\@centering&\llap{##}\tabskip\z@\cr}

\def\endeqnsystem{\@@eqncr\egroup$$\global\@ignoretrue} \makeatother

\def\SU{{\rm SU}}

\def\circa#1{\,\raise.3ex\hbox{$#1$\kern-.75em\lower1ex\hbox{$\sim$}}\,}

\usepackage{multicol}
\usepackage{color}
\definecolor{rosso}{cmyk}{0,1,1,0.4}
\definecolor{rossos}{cmyk}{0,1,1,0.55}
\definecolor{rossoc}{cmyk}{0,1,1,0.2}
\definecolor{blu}{cmyk}{1,1,0,0.3}
\definecolor{blus}{cmyk}{1,1,0,0.6}
\definecolor{bluc}{cmyk}{1,1,0,0.1}
\definecolor{verde}{cmyk}{0.92,0,0.59,0.25}
\definecolor{verdec}{cmyk}{0.92,0,0.59,0.15}
\definecolor{verdes}{cmyk}{0.92,0,0.59,0.4}
\definecolor{grigio}{cmyk}{0,0,0,0.07}
\definecolor{rosa}{cmyk}{0,0.1,0.1,0.02}
\definecolor{rosino}{cmyk}{0,0.05,0.05,0.02}
\definecolor{rosas}{cmyk}{0,0.3,0.25,0.05}
\definecolor{celeste}{cmyk}{0.1,0,0,0.02}
\definecolor{giallino}{cmyk}{0,0,0.4,0.02}
\definecolor{rosso}{cmyk}{0,1,1,0.4}
\definecolor{rossos}{cmyk}{0,1,1,0.55}
\definecolor{rossoc}{cmyk}{0,1,1,0.2}
\definecolor{blu}{cmyk}{1,1,0,0.3}
\definecolor{bluc}{cmyk}{1,1,0,0.1}
\definecolor{blucc}{cmyk}{0.7,0.5,0,0}
\definecolor{viola}{cmyk}{0,1,0,0.6}
\definecolor{viola2}{cmyk}{0,1,0.2,0.6}
\definecolor{verde}{cmyk}{0.92,0,0.59,0.25}
\definecolor{verdec}{cmyk}{0.92,0,0.59,0.15}
\definecolor{verdes}{cmyk}{0.92,0,0.59,0.4}
\definecolor{verdino}{cmyk}{0.12,0,0.09,0.05}
\definecolor{giallo}{cmyk}{0,0,1,0}
\definecolor{gialloverde}{cmyk}{0.44,0,0.74,0}

\usepackage[normalem]{ulem}

\font\tenrsfs=rsfs10 at 12pt

\font\sevenrsfs=rsfs7
\font\fiversfs=rsfs5
\newfam\rsfsfam
\textfont\rsfsfam=\tenrsfs
\scriptfont\rsfsfam=\sevenrsfs
\scriptscriptfont\rsfsfam=\fiversfs
\def\mathscr#1{{\fam\rsfsfam\relax#1}}
\def\Lag{\mathscr{L}}
\def\Lag{{\cal L}}

\def\eq#1{eq.~(\ref{#1})}
\def\beq{\begin{equation}}
\def\eeq{\end{equation}}
\def\bea{\begin{eqnarray}}
\def\eea{\end{eqnarray}}

\begin{document}\hfill
 %IFUP-TH/2011-XX\hfill 
  \centerline{CERN-PH-TH/2013-052}

\color{black}
\vspace{1cm}
\begin{center}
{\Huge\bf\color{magenta} The universal Higgs fit }\\
\bigskip\color{black}\vspace{0.6cm}{
{\large\bf Pier Paolo Giardino$^{a,b}$, Kristjan Kannike$^{{c,d}}$, \\[3mm]
Isabella Masina$^{e,f}$,
Martti Raidal$^{d,g}$
 {\rm and} Alessandro Strumia$^{a,d}$}
} \\[7mm]
{\it (a) Dipartimento di Fisica, Universit{\`a} di Pisa and INFN, Italy}\\[1mm]
{\it (b) CERN, Theory Division, CH-1211 Geneva 23, Switzerland}\\[1mm]
{\it  (c) Scuola Normale Superiore and INFN, Piazza dei Cavalieri 7, 56126 Pisa, Italy}\\[1mm]
{\it  (d) National Institute of Chemical Physics and Biophysics, R\"{a}vala 10, Tallinn, Estonia}\\[1mm]
{\it (e) Dipartimento di Fisica e Scienze della Terra dell'Universit\`a di Ferrara and INFN, Italy }\\[1mm]
{\it (f) CP$^\mathbf 3$-Origins and DIAS, Southern Denmark University, Denmark}\\[1mm]
{\it  (g) Institute of Physics, University of Tartu, Estonia}\\[3mm]
\end{center}
\bigskip
\bigskip
\bigskip

\centerline{\large\bf\color{blus} Abstract}

\begin{quote}\large
%We condense Higgs data into a `universal' form
%that allows to easily test any model.
%This is applied to various analyses: we extract from data
%the Higgs branching ratios, production cross sections, couplings,
%we consider composite Higgs, extra particles in the loops,
%anomalous top couplings, invisible decay width, 
%connection with Dark Matter models.
%Best fit regions lie around the Standard Model predictions and our well approximated by our `universal' fit.
%In particular we find that present data exclude the dilaton alternative to the Higgs and
%we derive the first measurement of the SM Higgs mass from the rates
%(rather than from the peak positions): $M_h = 124.3\pm1.9\GeV$.

We perform a state-of-the-art global fit to all Higgs data. 
We synthesise  them  into a  `universal'  form, which allows to
easily test any desired model.
We apply the proposed methodology to 
extract from data the Higgs branching ratios, production cross sections, couplings
and to analyse composite Higgs models, models with extra Higgs doublets, 
supersymmetry, extra particles in the loops, anomalous top couplings, and invisible Higgs decays into Dark Matter.  
Best fit regions lie around the Standard Model predictions and are well approximated by our `universal' fit.
Latest data exclude the dilaton as an alternative to the Higgs, and
disfavour fits with  negative Yukawa couplings.
We derive for the first time the SM Higgs boson mass from the measured rates,
rather than from the peak positions, obtaining $M_h = 125.0\pm 1.8\GeV$.

\end{quote}
%\end{abstract}

\thispagestyle{empty}

\newpage

%\tableofcontents

\section{Introduction}
\label{sec:intro}

 \medskip
After the discovery of a new  particle around 125.5 GeV announced 
by the ATLAS~\cite{ATLAS} and CMS~\cite{CMS} LHC collaborations during 2012,
all LHC and Tevatron collaborations  presented at the Moriond 2013 conference
their new results based on the full collected data.
These include the most important $\gamma\gamma,$ $ZZ^*$ and $WW^*$ channels as well as updates to the fermionic channels.
% In addition, Tevatron presented their final results for Higgs boson searches in the same conference. 
Such results will stay with us for next two years until LHC with full energy starts operating. Therefore it is 
the right moment to analyse their implications. 
%of 
% those experimental measurements on the SM Higgs boson as well as on models beyond the SM.

\smallskip

We  want to know if the new particle is the long-waited
Standard Model (SM) Higgs boson \cite{Englert:1964et,Higgs:1964ia,Higgs:1964pj,Guralnik:1964eu}. 
On one side, the experimental collaborations are measuring its discrete quantum numbers to check if it is a scalar.
On the other side, various theoretical groups~\cite{hits} started to approximatively
reconstruct from data its production cross section and its decay modes and consequently its couplings
%(into $\gamma\gamma$, $ZZ^*$, $WW^*$,  $\tau^+\tau^-$ and $b\bar{b}$)
to check if they agree with the SM predictions or with other models beyond the SM.
Clearly, this is a more significant test that can be precisely done only by the
experimental collaborations, which indeed started to present analyses along these lines.
However these experimental fits, presented in the form of likelihood plots within a few specific 
beyond-the-SM models,
are of little use to theorists who are interested in different models.

\smallskip

We here propose how experimental collaborations could report their
results in a model-independent and useful way, such that these results would be readily and reliably used by theorists
who want to test any desired model.
The  new ingredient that we introduce and that allows for this simplification is the assumption  that new physics can be approximated as a first-order perturbation with respect to the SM predictions. 
We find that this assumption is increasingly supported by measurements, 
that agree with the SM with precisions around the  20\% level.

\smallskip

Such results, obtained after two years of LHC operation
and with only  25/fb data per experiment, implies severe constraints on models where the Higgs boson is a portal 
to new physics. We analyse several models and rule out  alternative scenarios to the Higgs boson.

\smallskip\smallskip

The paper is organised as follows.
In section~\ref{sec:data} we present the data and our fitting procedure.
In section~\ref{sec:Mh} we derive the first measurement of the Higgs mass from the rates, rather than
from the position of the peaks in the $\gamma\gamma$ and $ZZ$ invariant mass distributions.
In section~\ref{sec:fit} we present the `universal' format for data  mentioned above.
Next, in section~\ref{sec:BSM} we present fits in various specific models, updating our previous results \cite{Giardino:2012ww,Giardino:2012dp} and comparing 
the full fit to the simplified `universal' fit to verify that it is a good approximation.
We fit Higgs cross sections in section~\ref{Hsigma}, Higgs couplings in~\ref{Hc},
composite Higgs models in~\ref{composite}, new physics in loops in~\ref{loops},
two Higgs doublet models in~\ref{sec:2HDM}, 
the MSSM in~\ref{MSSM},
the dilaton in~\ref{sec:dilaton},
the Higgs invisible width in~\ref{inv} and models where DM couples to the Higgs in~\ref{sec:DM}.
In section~\ref{sec:concl} we summarise the results and draw our conclusions.
%discuss ways to reconstruct the Higgs boson properties:
%1) mass; 2) branching ratio; 3) invisible width; 4) couplings.

% and present a concise 
%summary of all available Higgs data. 
%(for more technical details of our fitting procedure and 
%motivations of the scenarios we consider we refer the gentle reader to our previous paper~\cite{Giardino:2012ww}). 
%In addition, we perform here for the first time a generic Higgs fit based on the Gaussian approximation. 
%We show that the latter method, is already as good as the standard one, although being much simpler.

%In Sectionwe perform a generic Higgs fit to the data, using both the standard method and the Gaussian approximation.
%We show that the Gaussian approximation can already be sensibly applied.

%%%%%%%%%%%%%%%%%%%%%%%%%%
\section{The data}
\label{sec:data}

Searches for the SM Higgs boson have been carried out in proton-proton collisions at $\sqrt{s} = 7$ (2011 data) and $8$ TeV (2012 data)
with about 25/fb of total integrated luminosity at the LHC and in proton-antiproton collisions at $\sqrt{s} = 1.96$~TeV at the Tevatron.

%the analyzed data correspond to integrated luminosities of up to $5.1$ fb$^{-1}$ and $19.6$ fb$^{-1}$, respectively;
%for ATLAS, ...

There are four main production modes for Higgs boson from $ pp$ collisions. 
The gluon-gluon fusion production mode has the largest cross section, 
followed in turn by vector boson fusion (VBF), 
associated $Wh$ and $Zh$ production, 
and production in association with top quarks, $t\bar t h$. 
The cross sections for the Higgs boson production modes and
the decay branching fractions, together with their uncertainties, are taken from \cite{HXSWG}.  

%The relevant decay modes of the SM Higgs boson depend strongly on its mass $M_h$.
Our updated analysis uses the new data presented at the %HCP2012 and 
Moriond 2013 conference by the CMS, ATLAS and Tevatron collaborations \cite{cms:2013:bosonic, atlas:2013:bosonic, fermions:2013, Tevatron:2013}
in the following five decay modes: 
$\gamma\gamma$ \cite{gammagamma:2013}, 
$ZZ^*$ (followed by $ZZ^*$ decays to $4 \ell, 2\ell 2 \nu, 2 \ell 2 q, 2 \ell 2 \tau$) \cite{ZZ:2013}, 
$WW^*$ (followed by $WW^*$ decays to $\ell \nu \ell \nu,  \ell \nu q q$) \cite{WW:2013,atlas:2013:bosonic},  
$\tau^+\tau^-$ (followed by leptonic and hadronic decays of the $\tau$-leptons) \cite{tautau:2013} (we include the CMS $\tau^{+}\tau^{-}$ results updated at the end of 2013 \cite{tautau:cms:update:2013})
and $b\bar{b}$ \cite{Tevatron:2013,bb:2013} (the ATLAS $b\bar{b}$ result was updated at the EPS HEP 2013 \cite{bb:2013:epshep}), and the first tentative measurements 
in the $\mu^{+} \mu^{-}$ \cite{mumu:2013}, $Z\gamma$ \cite{Zgamma:2013} and $WWW$ \cite{WWW:2013}  channels, 
as well as their combination \cite{combination:2013}. We also include the $t\bar t h$ rate presented by ATLAS at the Moriond 2014 conference~\cite{tth}. Our latest analysis includes the ATLAS $\gamma\gamma$ \cite{atlas:gammagamma:ichep:2014}, $Z Z^{*}$ \cite{atlas:ZZ:ichep:2014}, $\mu^{+}\mu^{-}$ \cite{atlas:mumu:2014} and $t\bar{t}h$ \cite{atlas:ttH:ichep:2014}  and CMS $\gamma\gamma$ \cite{cms:gammagamma:ichep:2014}, $Z Z^{*}$ \cite{cms:ZZ:ichep:2014}, $WW^{*}$ \cite{cms:WW:ichep:2014}, $\tau^{+}\tau^{-}$ and $\mu^{+}\mu^{-}$ \cite{cms:fermions:ichep:2014} results presented at the ICHEP 2014 conference and at a seminar at CERN in July 2014 \cite{cms:higgs:cern:2014}.
Here and throughout, $\ell$ stands for electrons or muons and $q$ for quarks.

%[taken from CMS PAS 045]
For a given Higgs boson mass, the search sensitivity depends on the production
cross section of the Higgs boson, its decay branching fraction into the chosen final state, the
signal selection efficiency, the mass resolution, and the level of standard model backgrounds
in the same or a similar final state. 
For low values of the Higgs boson mass, the $h \to \gamma \gamma$ and $h \to ZZ^* \to 4 \ell$ channels play a special role 
due to the excellent mass resolution for the reconstructed diphoton and four-lepton final states, respectively. 
The $h \to  WW^* \to  \ell \nu \ell \nu$ channel provides high sensitivity but has relatively poor mass resolution due to the presence 
of neutrinos in the final state. 
The sensitivity in the $b \bar b$ and $\tau^+ \tau^-$ decay modes is reduced due to the large backgrounds and poor mass resolutions.

We include in our data-set all exclusive $\gamma\gamma$ and $\tau\tau$ sub-categories
described by the experimental collaborations by telling how much each Higgs production channel in the SM
contributes to the various rates.  Such information is fully included in our analysis.
We adopt the latest $\gamma\gamma$ data MultiVariate Analysis (MVA) from CMS. The two CMS $\gamma\gamma$ analyses (cut-based and MVA) show different signal rates (compatible within $1 \sigma$), and the latter one is closer to the SM. We combine all experiments finding an average $\gamma\gamma$ rate very close to the SM prediction. Consequently our results differ from previous analyses~\cite{hits}
performed without including the latest CMS $\gamma\gamma$ data.

This is an important issue because,
while most of the presented LHC results are consistent with the SM predictions within experimental errors,
there are a few unexpected new developments that warrant additional discussion.  
The most important of them is the discrepancy between the ATLAS and CMS results in the $h\to\gamma\gamma$ channels. 
With full integrated luminosity datasets, ATLAS finds an overall rate of $1.29\pm 0.30$ and
 CMS finds $1.13\pm0.24$.
%The two measurements are compatible within $2\sigma$.
 Finally, the two Higgs boson
 mass determinations in ATLAS, from the peaks in the $\gamma\gamma$ and $ZZ$ channels, differ by $2\sigma$. 
 Both experiments have cross checked their 
 analyses and reached conclusions  that these deviations appear to be due to statistical fluctuations of both signal and background.
 This conclusion implies that: $(i)$~combining all data in a global fit is meaningful and increases the precision; 
 $(ii)$~selecting instead any single measurement, for example the ATLAS excess in $\gamma\gamma$,
is not  justified and would introduce a bias in the data. 
%$(iii)$~in order to test new
% physics models against the Higgs boson data one should study deviations from the  global fit.

\begin{figure}[t]
 $$ \includegraphics[height=8.3cm]{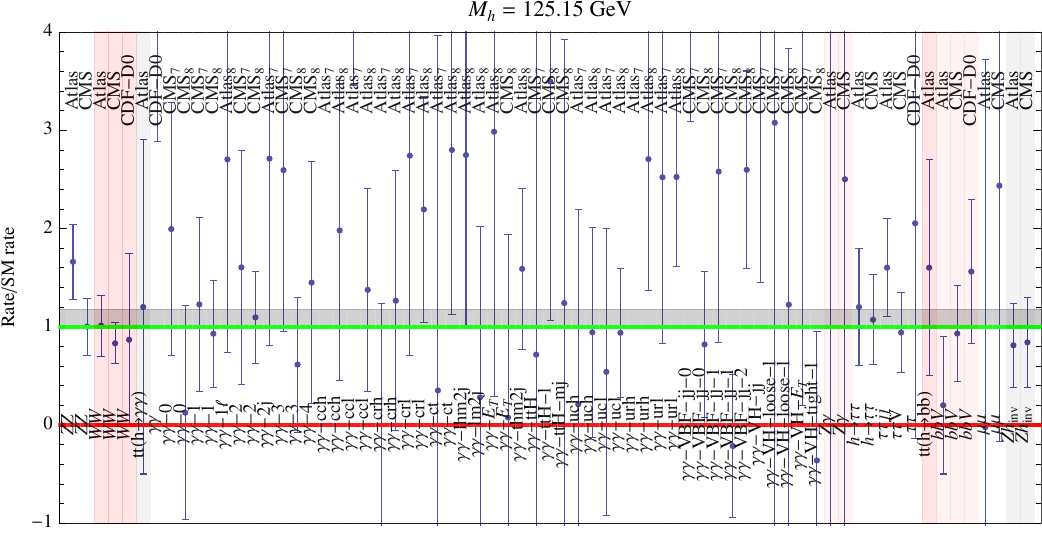} $$
\caption{\em  
Measured Higgs boson rates at ATLAS, CMS, CDF, D0 and their average (horizontal gray band at $\pm1\sigma$).
Here 0 (red line) corresponds to no Higgs boson, 1 (green line) to the SM Higgs boson
(including the latest data point, which describes the invisible Higgs rate).
\label{fig:data}}
\end{figure}

%by the CMS collaboration in the $h \to ZZ^* \to 4 \ell$ \cite{ZZ:new}, 
%and by CMS and ATLAS collaborations in $h \to WW^* \to 2 \ell 2 \nu$ \cite{WW:new}, $h \to b\bar{b}$ \cite{bb:new}, and $h \to \tau^+ \tau^-$ 
%\cite{tautau:new} channels, 
%and as their combination \cite{comb:new}. 
%The $h \to \gamma\gamma$ channel saw no update from the data presented in July 2012 \cite{gammagamma};
%however, we now take into account the various $\gamma\gamma$ subchannels of the CMS and ATLAS. ATLAS also did not update their $ZZ^*$ channel
%data from that shown in July \cite{ZZ}. The $h \to \tau^+ \tau^-$ channel data was presented for the first time.

%\subsection{The fitting procedure}
\label{stat}

\bigskip

The experimental collaborations report Higgs boson rates $R$
in units of the central value of the SM prediction.
Their results could be fully encoded in a likelihood ${\cal L}(R,M_h)$, 
but only a limited amount of information is reported by the experiments.
Often the experimental collaborations report the measured rates as $R^{\rm exp}\pm R^{\rm err}$:
we use the results in this form whenever available.
Sometimes collaborations only report the upper bounds on rates at 95\% C.L., $R^{\rm limit}_{\rm observed}$, 
and the expected upper bound at $95\%$ C.L. in absence of a Higgs boson signal, $R^{\rm limit}_{\rm expected}$,
as function of the Higgs boson mass $m_h$. 
Assuming that the $\chi^2 = -2 \ln {\cal L}$ has a Gaussian form in $R$,
%$\chi^2 = (R-\mu)^2/\sigma^2$,
these two experimental informations allow one to extract the mean $R^{\rm exp}$ and the standard deviation $R^{\rm err}$ as
$R^{\rm exp} = R^{\rm limit}_{\rm observed} - R^{\rm limit}_{\rm expected}$ and
$R^{\rm err} = R^{\rm limit}_{\rm expected}/1.96$,
where $1.96$ arises because $95\%$ confidence level corresponds to about 2 standard deviations~\cite{Giardino:2012ww}.\footnote{A similar procedure
was described by Azatov et al.\ in~\cite{hits}.}
The $\chi^2$ is approximated as
\beq
\chi^2 =
\sum_I \frac{(R_I^{\rm exp}-1)^2}{(R_I^{\rm err})^2}, \eeq
where the sum runs over all measured Higgs boson rates $I$.

The theoretical uncertainties on the Higgs production cross sections $\sigma_j$ start to be non-negligible and affect the observed rates in a correlated way.%
\footnote{
Note that the size of the theory uncertainty depends on the applied cuts, and that the scaling of the production cross section to yield the best-fit value relative to the theoretical central value may
be different for very different sets of selection cuts (due to the QCD and
PDF uncertainties).}
We take into account such correlations in the following way.
We subtract from the total uncertainty $R^{\rm err}_I$ the theoretical component due to
the uncertainty in the production cross sections, 
obtaining the purely experimental uncertainty, $R_I^{\rm{err-exp}}$. The
theoretical error is reinserted by defining a $\chi^2$ which depends on the production cross sections $\sigma_j$,
\beq
\chi^2 =
\sum_I \frac{( R_I^{\rm exp}-R_I^{\rm th} (\sigma_j))^2}{(R_I^{\rm{err-exp}})^2} + 
\sum_j
\frac{(\sigma_j - \sigma_j^{\rm th})^2}{(\sigma_j^{\rm err})^2},
\eeq
and marginalising it with respect to the free parameters $\sigma_j$, 
constrained to have a central value $\sigma_j^{\rm th}$ and an uncertainty $\sigma_j^{\rm err}$ given e.g. at $\sqrt{s} = 8 \TeV$ by~\cite{HXSWG}
\beq \begin{array}{ll}
\sigma(pp\to h)_{\rm th} =(19.4\pm2.8)\pb,\qquad&
\sigma(pp\to jjh)_{\rm th}=(1.55\pm 0.04)\pb,\\
\sigma(pp\to Wh)_{\rm th}=(0.68\pm0.03)\pb,\qquad&
\sigma(pp\to Zh)_{\rm th} = (0.39\pm0.02)\pb ,\\
\sigma(pp\to t\bar t h)_{\rm th} = (0.128\pm0.018)\pb. & 
\end{array}\eeq
See also \cite{Duhrssen:2004cv}. We neglect the relatively small uncertainties on the SM theoretical predictions for
Higgs branching ratios,
dominated by a $4\%$ uncertainty on the $h\to b\bar b$ width.

We summarise all data 
in fig.\fig{data} together with their  $1\sigma$ error-bars.
%, as derived by usfollowing the above-described statistical procedure. 
%The horizontal green line is the SM prediction, and the horizontal red line is the background-only rate expected in the absence of a Higgs boson.
The grey band shows the $\pm1\sigma$ range for the naive weighted average of all rates: $1.08\pm 0.09$.
%\footnote{This is obtained by performing a global fit, where we rescale all Higgs rates by a common factor with respect to the SM prediction.  
%Given that we neglect correlations among experimental uncertainties, the result is simply the weighted average of all rates reported by the experiments.}
%\beq 
%\frac{\rm Measured~Higgs~rate}{\rm SM~prediction}= \frac{R_i^{\rm exp}}{R_i^{\rm th}(\sigma_j^{\rm exp})}
% \eeq
It lies along the SM prediction of 1 (horizontal green line) and, performing  a naive average with Gaussian errors,
it is almost 10
$\sigma$ away from 0 (the horizontal red line is the background-only rate expected in the absence of a Higgs boson). 

\begin{table}$$
\begin{array}{c|ccc}
\hbox{Higgs mass in GeV} & \hbox{from $h\to ZZ$} & \hbox{from $h\to \gamma\gamma$} & \hbox{$ZZ$ and $\gamma\gamma$} \\ \hline
%\hbox{From ATLAS} & 124.5\pm0.5_{\rm stat}\pm0.06_{\rm syst} &
%126.0\pm0.4_{\rm stat}\pm0.3_{\rm syst} & 125.36\pm0.37_{\rm stat}\pm0.18_{\rm syst}  \\
%\hbox{From CMS~~~} & 125.6\pm0.4_{\rm stat}\pm0.2_{\rm syst} &
%124.7\pm0.3_{\rm stat}\pm0.15_{\rm syst} & 125.03 \pm 0.27_{\rm stat}\pm0.14_{\rm syst}\\
%\hbox{ATLAS and CMS} & 124.64\pm0.28 & 125.02\pm0.27 & 125.15\pm0.24
\hbox{From ATLAS} & 124.5\pm0.5\pm0.06 &
126.0\pm0.4\pm0.3 & 125.36\pm0.37\pm0.18  \\
\hbox{From CMS~~~} & 125.6\pm0.4\pm0.2 &
124.7\pm0.3\pm0.15 & 125.03 \pm 0.27\pm0.14\\
\hbox{ATLAS and CMS} & 124.64\pm0.28 & 125.02\pm0.27 & 125.15\pm0.24
\end{array}
$$
\caption{\label{tab:Mh}\em 
Determinations of the Higgs mass from the peaks of the invariant mass
of $\gamma\gamma$ and $ZZ$ events, taking into account the latest CMS and ATLAS results presented at the ICHEP 2014.  The first uncertainty is statistical and the second is systematical.}
\end{table}

\section{Reconstructing the Higgs mass}\label{sec:Mh}
The CMS and ATLAS collaborations reported
measurements of the pole Higgs mass $M_h$ obtained as the position of the
peaks observed in the invariant mass of the $h\to \gamma\gamma$ and $h\to ZZ\to 4\ell$ distributions.
Averaging the results summarised in table~\ref{tab:Mh} we find
\beq M_h = 125.15\pm0.24~\GeV
 \qquad\hbox{(Higgs mass extracted from the $ZZ$ and $\gamma\gamma$ peaks).}
%=
%\left\{\begin{array}{ll}
%124.7\pm0.3_{\rm stat}\pm0.15_{\rm syst}~\GeV & \hbox{CMS $\gamma\gamma$}\\
%125.6\pm0.4_{\rm stat}\pm0.2_{\rm syst}~\GeV & \hbox{CMS $ZZ$}\\
%126.0\pm0.4_{\rm stat}\pm0.3_{\rm syst}~\GeV & \hbox{ATLAS $\gamma\gamma$}\\
%124.5\pm0.5_{\rm stat}\pm0.06_{\rm syst}~\GeV & \hbox{ATLAS $ZZ$}\\
%\end{array}\right. \ .
\eeq
The measurements are mutually compatible, and the uncertainty is small enough that  
in the subsequent fits to rates  we can fix $M_h$ to its combined best-fit value. 
We combined all uncertainties in quadrature, using the standard Gaussian error propagation and
neglecting correlations among systematic uncertainties.
The averages within each experiment agree with those reported by the experiments.
%The ATLAS collaboration reports the combined value for the Higgs mass, based on the $\gamma\gamma$ and $ZZ$ channels, as 
%$M_h = 125.5 \pm 0.2_{\rm stat} {}_{-0.6}^{+0.5}{}_{\rm syst}$ (best fit signal strength $R = 1.43 \pm 0.16_{\rm stat} \pm 0.14_{\rm syst}$) \cite{atlas:mass:2013}, whereas CMS gives $M_h = 125.7 \pm 0.3_{\rm stat} \pm 0.3_{\rm syst}$ based on $\gamma\gamma$, $ZZ$, $WW$, $\tau\tau$ and $bb$ (best fit signal strength $R = 0.80 \pm 0.14$) \cite{cms:mass:2013}.
%Averaging these mass values reproduces the $M_h$ of \eq{eq:Mhpeak}.

\medskip

We here discuss how the Higgs mass can be independently measured, with a larger uncertainty,
 by requiring that the measured rates agree with their 
SM predictions within their uncertainties.
Such predictions have a dependence on the Higgs mass that, around 125 GeV,  can be approximated as
\beq \sigma(pp\to X) \approx \sigma(pp\to X)_{M_h = 125\GeV} \times [1 + c_X \times (M_h - 125\GeV)].\eeq
In table~\ref{tab:Mhrate}  we list the values of the coefficients $c_X$ and of the measured rates for the various processes
averaging all experiments,
as well as the  Higgs mass indirectly derived from such rates.  
We see that the single best indirect determination of $M_h$ comes from the $h\to WW$ rates, 
that presently have no sensitivity to $M_h$ if one wants to measure it from a mass peak.
We see that best indirect determinations of $M_h$ comes from the $h\to WW$ rates (which presently have no sensitivity to $M_h$ if one wants to measure it from a mass peak) and from $h\to ZZ$.
On the other hand, the $h\to\gamma\gamma$ signal that offers the best peak measurement of $M_h$
has very little indirect sensitivity to $M_h$, because the $\gamma\gamma$ rate happens to have a weak dependence on $M_h$.
Averaging over all channels we find 
\beq M_h = 125.0\pm 1.8~\GeV \qquad\hbox{(Higgs mass extracted from the rates, assuming the SM)}\eeq
which is compatible with the determination of the pole Higgs mass obtained in a model-independent way
from the positions of the peaks.

\begin{table}
$$ \begin{array}{c|cccccc}
\hbox{Process $X$} & 
h \to WW & h\to ZZ  & 
h\to \gamma\gamma & 
Vh\to Vbb & 
h\to\tau\tau
\\ \hline
\hbox{Sensitivity $c_X$} 
& 6.4\%/\GeV
& 7.8\%/\GeV
& -1.5\%/\GeV
& -5.4\%/\GeV
& -4.1\%/\GeV\\
\hbox{Measured rate/SM} 
& 0.89\pm0.17 
& 1.24\pm0.23
&1.27\pm0.17 
&0.96\pm0.34
&1.18\pm0.24\\
\hbox{Higgs mass in GeV}
& 123.7\pm 2.7
& 128.6\pm 3.0
& 128\pm 11
& 126\pm 6
& 121\pm 6
\end{array}$$
\caption{\label{tab:Mhrate}\em Determinations of the Higgs mass from the measured Higgs rates, assuming the SM predictions for such rates. We do not use here the independent determination of the Higgs mass from the peak positions in the $\gamma\gamma$ and $ZZ$ energy spectra.}
\end{table}

%Presently  only the $h\to WW$ rates have been measured precisely enough to provide significant indirect measurements
%of the Higgs mass:
%\beq M_h = \left\{\begin{array}{ll}
%129.5\pm 6.4 & \hbox{ATLAS}\\
%121.5\pm 3.4 & \hbox{CMS}
%\end{array}\right.\eeq

\section{The universal Higgs fit}\label{sec:fit}
We perform the most generic fit
in terms of a particle $h$ with couplings 
to pairs of $t,b,\tau,W,Z,g,\gamma$ equal to
$r_t,r_b, r_\tau,r_W,r_Z,r_g,r_\gamma$ in units of the SM Higgs coupling.%

\footnote{
The $r_{i}$ are equivalent to the $\kappa_{i}$ parametrisation as defined in \cite{LHCHiggsCrossSectionWorkingGroup:2012nn}.}
This means, for example, that the coupling to the top is given by $ r_t (m_t/V) h\bar t t$,
where $r_t = 1$ in the SM and $V=246\GeV$ (from the measurement of the Fermi constant \cite{Tishchenko:2012ie}) is the Higgs vacuum expectation value.
Similarly, the $h\gamma\gamma$ coupling is assumed to be $r_\gamma$ times its SM prediction.
In the SM this couplings first arises at one loop level.
Experiments are starting to probe also the $h\bar\mu\mu$ and the $hZ\gamma$ effective couplings,
so that also the corresponding $r_\mu$ and $r_{Z\gamma}$ parameters will start to be measured.
This discussion can be summarized by the following effective Lagrangian: 
\begin{eqnarray} 
\Lag_h &=&
r_t \frac{m_t}{V} h \bar tt + 
r_b \frac{m_b}{V} h \bar bb + 
r_\tau \frac{m_\tau}{V} h \bar\tau \tau + 
r_\mu \frac{m_\tau}{V} h \bar\mu\mu + 
r_Z \frac{M_Z^2}{V} h Z_\mu^2 + 
r_W \frac{2M_W^2}{V} h W^+_\mu W^-_\mu + \nonumber \\
&&+ r_\gamma   c_{\rm SM}^{\gamma\gamma}  \frac{\alpha}{ \pi V} h F_{\mu\nu}F_{\mu\nu} 
+ r_g   c_{\rm SM}^{g g}  \frac{\alpha_s}{12 \pi V} h G^a_{\mu\nu} G^a_{\mu\nu} 
+ r_{Z\gamma}   c_{\rm SM}^{Z\gamma}  \frac{\alpha}{ \pi V} h F_{\mu\nu}Z_{\mu\nu} .
\label{Lh}
\end{eqnarray}
The various SM loop coefficients $c_{\rm SM}$ are summarised in appendix~\ref{app:L}.
This Lagrangian is often written in a less intuitive but practically equivalent form by either
using $\SU(2)_L\otimes{\rm U}(1)_Y$-invariant effective operators, or 
assuming that  the Higgs is the pseudo-Goldstone boson of a spontaneously broken global symmetry
and writing its chiral effective theory~\cite{hits}.
We do not consider a modified Higgs coupling to charm quarks, given that  $h\to c\bar c$ decays at LHC
are hidden by the QCD background.  While we cannot exclude that new physics affects $h\to c\bar c$ much more than all other Higgs properties,
for simplicity we proceed by discarding this possibility.

\medskip

Furthermore, we take into account the possibility of Higgs decays into invisible particles $X$, such as Dark Matter or neutrinos~\cite{Khlopv},
with branching ratio ${\rm BR}_{\rm inv}$.
In almost all measured rates (with the exception of the last data-point in fig.\fig{data}:
the direct measurement of the invisible Higgs width)
${\rm BR}_{\rm inv}$ is equivalent to a common reduction $r$ of
all the other Higgs couplings, ${\rm BR}_{\rm inv} \simeq 1 - r^2$, such that
${\rm BR}_{\rm inv}$ is indirectly probed by data~\cite{Giardino:2012ww}.
The only observable that directly probes an invisible Higgs width is the
$pp \to Z h \to \ell^+\ell^-~ \bar X X$ rate measured by ATLAS~\cite{ATLASinv} and CMS~\cite{CMSinv}, which implies 
\beq \hbox{BR}_{\rm inv} = -0.18\pm 0.31.\eeq
%<65% (measured); <84% (expected)

Any possible new-physics model can be described as specific values of the $r_i$ parameters.
Several examples are provided in section~\ref{sec:BSM}.

Following the procedure described in the previous section,
we approximatively extract from data the function  
\beq \chi^2(r_t,r_b,r_\tau,r_W,r_Z,r_g,r_\gamma, r_{Z\gamma}, r_\mu,{\rm BR}_{\rm inv}),
\label{eq:chi2universal}\eeq
which  describes all the information contained in Higgs data.
We find $\chi^2 = 59.8$ at the best fit (69 data points, 10 free parameters), marginally better than the SM fit, $\chi^2_{\rm SM}=66.2$ (no free parameters).

%Especially after taking into account the theoretical uncertainties on the production cross sections,
%\xxx{Alessandro: do present exp data already take these uncertainties into account or not?  This MUST be clarified. Isa: Yes, I think so, see the discussion
%in the webpage of the Higgs Xsection WG}

\begin{figure}[t]
 $$ \includegraphics[width=\textwidth]{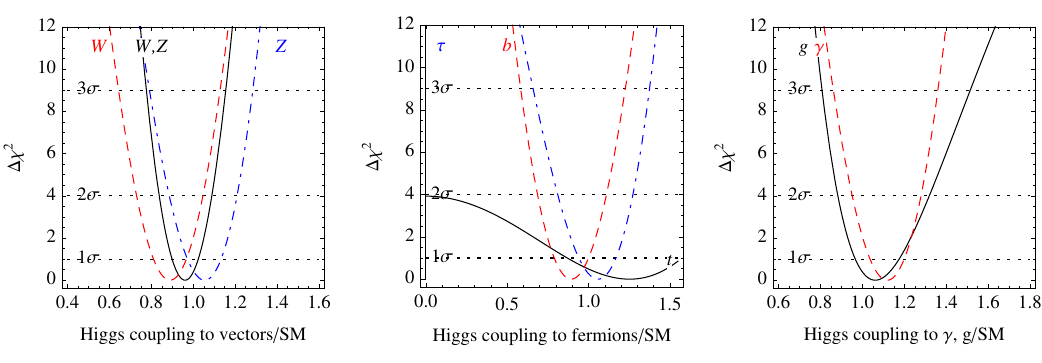}$$ 
\caption{\em 
$\chi^2$ as function of the model-independent Higgs couplings $r_i$ to the various SM particles,
varying them one-by-one (the others are set to unity).
\label{fig:g}}
\end{figure}

\subsection{Universal fit to small new physics effects}
The universal $\chi^2$ of \eq{eq:chi2universal}
has a too complicated form to be reported analytically,
and depends on too many variables to be reported in numerical form, such as plots or tables.
%Since the $\chi^2$ depends on many parameters, its also impossible to report it as plots or numerical tables.
For these reasons, previous analyses~\cite{hits,Giardino:2012ww,Giardino:2012dp} focused on particular {\rm BSM} models with a reduced number of parameters.
For example, fig.\fig{g} shows the fit as function of each $r_i$, setting all others to their SM values of unity:
we see that the $\chi^2$ are approximately parabolic. 

We here observe that Higgs
data are converging towards the SM predictions with small errors, thereby it is time to start making the approximation 
\beq r_i = 1+\epsilon_i\qquad\hbox{with} \qquad \epsilon_i \text{ small} % \ll1 
\eeq
and BR$_{\rm inv}=\epsilon_{\rm inv}$.
The observable rates $R_I$ are computed at first order in $\epsilon_i$, and consequently
the  $\chi^2$ is expanded up to second order in $\epsilon_i$.
As well known, this Gaussian approximation is a great simplification; for example
marginalisations over nuisance parameters just becomes minimisation, which preserves the Gaussian form.
Fig.\fig{g} suggests that this approximation already seems reasonably good, particularly
in the range of 1 or 2 standard deviations form the central value.

%This approximation means that each observable rate $R_I$ is computed at first order in $\epsilon_i$.
For LHC at $8\TeV$  the main observables are approximated as
\beq\begin{array}{rcl} 
R_{h\to W W} &=& 1-1.14 \epsilon _b+1.58 \epsilon _g- \epsilon _{\text{inv}}-0.04 \epsilon _t+1.72 \epsilon _W+0.02 \epsilon _Z-0.13 \epsilon _{\tau }  \\ 
R_{h\to Z Z} &=& 1-1.14 \epsilon _b+1.58 \epsilon _g- \epsilon _{\text{inv}}-0.04 \epsilon _t-0.28 \epsilon _W+2.02 \epsilon _Z-0.13 \epsilon _{\tau }  \\ 
R_{h\to \tau  \tau } &=& 1-1.14 \epsilon _b+1.58 \epsilon _g- \epsilon _{\text{inv}}-0.04 \epsilon _t-0.28 \epsilon _W+0.02 \epsilon _Z+1.87 \epsilon _{\tau }  \\ 
R_{h\to \gamma  \gamma } &=&1 -1.14 \epsilon _b+1.58 \epsilon _g- \epsilon _{\text{inv}}-0.04 \epsilon _t-0.45 \epsilon _W-0.06 \epsilon _Z-0.13 \epsilon _{\tau }+2 \epsilon _{\gamma }  \\ 
R_{h\to  b b} &=& 1+0.86 \epsilon _b+1.58 \epsilon _g- \epsilon _{\text{inv}}-0.04 \epsilon _t-0.28 \epsilon _W+0.02 \epsilon _Z-0.13 \epsilon _{\tau }  \\ 
R_{V(h\to  b b)} &=& 1+0.86 \epsilon _b-0.17 \epsilon _g- \epsilon _{\text{inv}}-0.05 \epsilon _t+0.83 \epsilon _W+0.67 \epsilon _Z-0.13 \epsilon _{\tau },
\end{array}\eeq
where these expressions have been obtained by performing a first-order Taylor expansion in all the $\epsilon$ parameters of
the full non-linear expressions.  For all observables but the last one, we have assumed the total Higgs production cross section.
When fitting the many real observables, we take into account the relative contribution of each production cross section, as determined
by experimental cuts.
For $h\to \gamma\gamma$ we here considered the gluon fusion production channel,
and this makes the coefficients of $\epsilon_{Z,W}$ somehow different from the other channels.
The full $\chi^2$ can now be reported in a simple form.
Indeed the  $\chi^2$ is a quadratic function of the $\epsilon_i$, and it is usually written as
\beq\label{eq:chiq}
\chi^2 =\sum_{i,j} (\epsilon_i - \mu_i) (\sigma^2)^{-1}_{ij}  (\epsilon_j - \mu_j),\qquad\hbox{where}
\qquad (\sigma^2)_{ij} = \sigma_i \rho_{ij} \sigma_j,\eeq
in terms of the mean values $\mu_i$ of each parameter $\epsilon_i$, 
of its error $\sigma_i$ and in terms of the correlation matrix
$\rho_{ij}$. 
We believe that this is the most useful form in which experimental collaborations could report their results.
From our approximated analysis of LHC 
 and Tevatron \cite{Tevatron:2013} data we obtain:
  \beq  \label{universal}
  \begin{array}{rcl} 
\epsilon _b &=& -0.19 \pm 0.28\\ 
\epsilon _g &=& -0.13 \pm 0.20\\ 
\epsilon _{\text{inv}} &=& -0.22 \pm 0.20\\ 
\epsilon _W &=& -0.20 \pm 0.13\\ 
\epsilon _Z &=&    \phantom{-}0.00 \pm 0.10\\ 
\epsilon _{\gamma } &=&    \phantom{-} 0.00 \pm 0.14\\ 
\epsilon _{\tau } &=& -0.03 \pm 0.17\\ 
\end{array}\qquad\rho = \left(
\begin{array}{ccccccc}
 1 & 0.70 & 0.04 & 0.52 & 0.38 & 0.58 & 0.59 \\
 0.70 & 1 & 0.43 & 0.38 & 0.11 & 0.40 & 0.52 \\
 0.04 & 0.43 & 1 & 0.46 & 0.13 & 0.40 & 0.34 \\
 0.52 & 0.38 & 0.46 & 1 & 0.44 & 0.63 & 0.45 \\
 0.38 & 0.11 & 0.13 & 0.44 & 1 & 0.42 & 0.33 \\
 0.58 & 0.40 & 0.40 & 0.63 & 0.42 & 1 & 0.54 \\
 0.59 & 0.52 & 0.34 & 0.45 & 0.33 & 0.54 & 1 \\
\end{array}
\right)
  \eeq
 We have not reported the central value of $r_t=1+\epsilon_t $, 
of $\epsilon_{Z\gamma}$ and of $\epsilon_\mu$ because they presently are known only up to
uncertainties much larger than 1.
Future searches for $t\bar t h$ production, for $h\to Z\gamma$ and for $h\to\mu^+\mu^-$ will 
improve the situation.

\smallskip

In many models  the Higgs couplings to vectors satisfy $\epsilon_W = \epsilon_Z$, because of
$\SU(2)_L$ invariance.
Furthermore,  in many models LEP precision data
force $\epsilon_W$ and $\epsilon_Z$ to be very close to 0.
This restriction can of course be implemented by just setting these parameters to be equal or vanishing
in the quadratic $\chi^2$.
%The result can be again represented in the form of \eq{eq:chiq} where now
%\beq
%\begin{array}{rcl} 
%\epsilon _b &=&      +0.00 \pm 0.29\\ 
%\epsilon _g &=& -0.15 \pm 0.23\\ 
%\epsilon _{\text{inv}} &=& -0.18 \pm 0.23\\ 
%\epsilon _{\gamma } &=& +0.06 \pm 0.11\\ 
%\epsilon _{\tau } &=& +0.08 \pm 0.15\\ 
%\end{array}\qquad\rho = \left(
%\begin{array}{ccccc}
% 1 & 0.70 & -0.23 & 0.33 & 0.23 \\
% 0.70 & 1 & 0.37 & 0.26 & 0.18 \\
% -0.23 & 0.37 & 1 & 0.19 & 0.09 \\
% 0.33 & 0.26 & 0.19 & 1 & 0.27 \\
% 0.23 & 0.18 & 0.09 & 0.27 & 1 \\
%\end{array}
%\right)\eeq
%where  we now introduced as explicit additional parameter the invisible branching ratio of the Higgs.

Since the uncertainties on the $\epsilon_i$ parameters are now smaller then 1, the universal approximation 
starts to be accurate.
In the next sections, where we analyze several specific models,  we will systematically compare 
our full numerical fit  (plotting best fit regions in yellow with continuous contours at the 90 and 99\% C.L.)
with the universal approximation (best fit ellipsoidal regions in gray with dotted contours, at the same confidence levels).

%%%%%%%%%%%%%%%%%%%%%%%%%%%%%%%%
\section{Model-dependent Higgs fits}
\label{res}\label{sec:BSM}

%\begin{figure}[t]
%$$\includegraphics[width=0.8\textwidth]{figs8TeV2/data2}$$
%\caption{\em Predictions for the Higgs boson rates in different scenarios: SM, free branching ratios of loop processes,
%free couplings, dilaton.
%\label{fig:data2}}
%\end{figure}

\begin{figure}[t]
$$\includegraphics[width=0.44\textwidth]{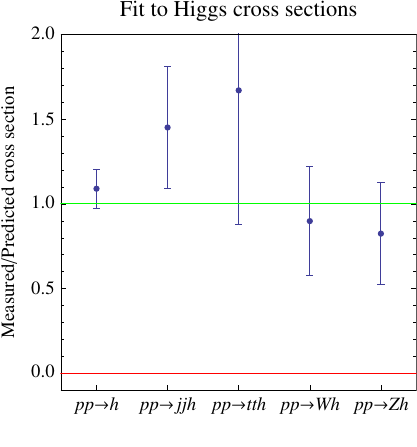}\qquad
\includegraphics[width=0.45\textwidth]{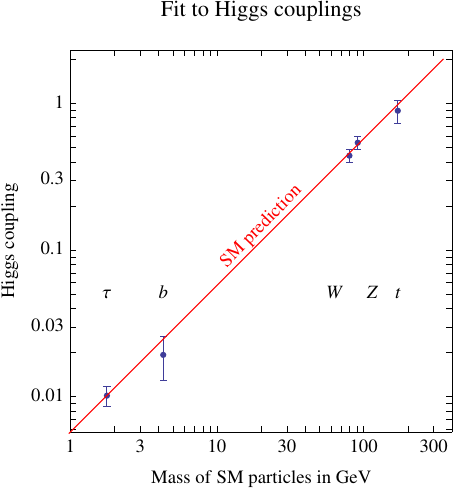}$$ 
\caption{\em {\bf Left}: reconstruction of the Higgs production cross sections in units of the SM prediction.
{\bf Right}:
reconstruction of the Higgs couplings to the $t,Z,W,b,\tau$, assuming
that no new particles exist. The SM predicts a proportionality between the Higgs couplings and the masses of the fermions and the squared masses of the vector bosons (diagonal line).
\label{fig:fitcouplings}}
\end{figure}

\subsection{Higgs production cross sections}\label{Hsigma}
Assuming the SM predictions for Higgs decay fractions, we extract from the data the 
Higgs production cross sections.  
Given that measured  rates of various exclusive and inclusive Higgs channels agree with their SM predictions,
we find that production cross sections also agree with SM predictions,
as shown in the left panel of fig.\fig{fitcouplings}.
As expected, the most precisely probed cross section is the
dominant one, $\sigma(pp\to h)$.
At the opposite extremum $\sigma(pp\to jjh)$ is still largely unknown.
The uncertainties on the reconstructed cross sections are correlated,
although we do not report the correlation matrix.

%\beq
%\begin{array}{rcl} 
%\sigma(pp\to h)/\sigma(pp\to h)&=& 0.95 \pm 0.12\\ 
%\sigma(pp\to jjh)&=& 1.40 \pm 0.42\\ 
%\sigma(pp\to V h) &=& 1.01 \pm 0.28\\ 
%\end{array}\qquad\rho = \left(
%\begin{array}{ccc}
% 1 & -0.29 & -0.12 \\
% -0.29 & 1 & 0 \\
% -0.12 & 0 & 1 \\
%\end{array}
%\right)
%\eeq

%
\begin{figure}[t]
$$\hspace{-0.01\textwidth}
\includegraphics[width=0.315\textwidth]{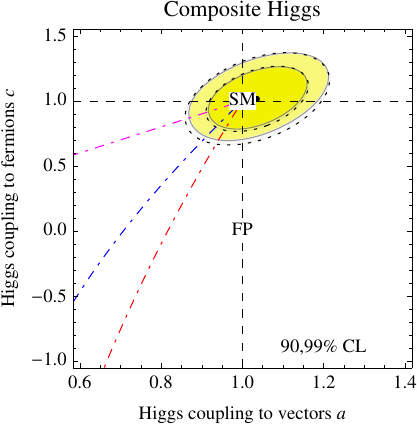}\quad
\includegraphics[width=0.315\textwidth]{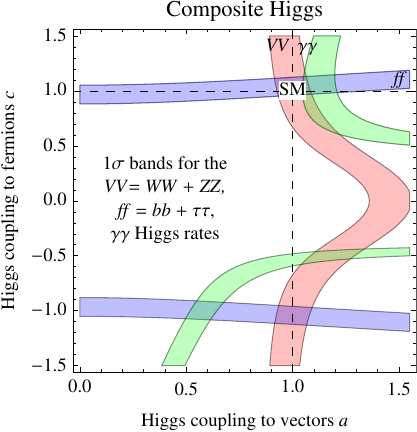}\quad
\includegraphics[width=0.315\textwidth]{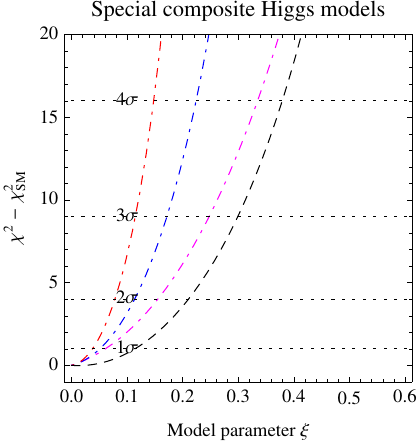}
$$ 
\caption{\em {\bf Left}: fit of the Higgs boson couplings
assuming common rescaling
factors $a$ and $c$ with respect to the SM prediction for couplings to
vector bosons and fermions,  respectively. 
The two sets of contour lines are our full fit (continuous) and our approximated `universal' fit (dotted).
{\bf Middle}: $1\sigma$ bands preferred by the three independent overall rates within the model.
{\bf Right}: 
values of the $\chi^2$ along the trajectories in the $(a,c)$ plane
shown in the left panel, and given by
$a=\sqrt{1-\xi}$ and
$c=a$ (magenta)
$c=(1-2\xi)/a$ (blue)
$c=(1-3\xi)/a$ (red),
as motivated by composite Higgs models~\cite{models}.
The black dashed curve corresponds to $a=1$ and $c=1-\xi$.
\label{fig:fitac}}
\end{figure}

\subsection{Higgs  couplings}\label{Hc}
We here extract from data the Higgs boson couplings to vectors and fermions, assuming 
that only the SM particles contribute to the $h\to gg,\gamma\gamma, \gamma Z$ loops.
This amounts to restricting the universal fit in terms of the $r_i$ parameters by setting
the parameters for loop couplings to 
\beq r_g = r_t,\qquad
r_\gamma %= \frac{ 4r_t A_f/3 + r_W A_W}{4 A_f/3 + A_W}
\approx 1.282 r_W - 0.282 r_t,\qquad
r_{Z\gamma}\approx 1.057 r_W - 0.057 r_t.
\label{eq:loops}
\eeq
These numerical expressions are obtained by rescaling 
the expressions for the SM loops summarised in appendix~\ref{app:L}.
In particular, the $W$ loop (rescaled by $r_W$)
and the top loop (rescaled by $r_t$)
contribute to $h\to\gamma\gamma$  with a negative interference.

Under this assumption the top coupling of the Higgs, $r_t$,  becomes indirectly probed via the loop effects.
The fit to the couplings is shown in fig.\fig{fitcouplings}b and agrees with the SM predictions
(diagonal line),
signalling that the new boson really is the Higgs.
The correlation matrix can be immediately obtained 
by inserting \eq{eq:loops} into the universal $\chi^2$ of \eq{eq:chiq}.

We allow the SM prediction to vary in position and slope by
assuming that the Higgs couplings to particles with mass $m$
are given by $(m/v')^p$.
Taking into account all correlations,
we find that data imply parameters $p$ and $v'$ 
 close to the SM prediction of $m/v$ (diagonal line in fig.\fig{fitcouplings}b):
\beq p=1.00\pm0.03,\qquad v' = v(0.97\pm0.06)\eeq 
with a $11\%$ correlation.

\subsection{Composite Higgs models}\label{composite}
Models where the Higgs is composite often assume the further restriction, in addition to \eq{eq:loops},
of a common 
rescaling with respect to their SM values
of the Higgs boson  couplings to the $W,Z$ bosons and a common rescaling of
the Higgs boson  couplings to all fermions.
These rescalings are usually denoted as $a$ and $c$, respectively:
\beq r_t = r_b = r_\tau =r_\mu= c,\qquad
r_W=r_Z = a.  \eeq
The resulting fit is shown in the left panel of fig.\fig{fitac}.
We see that  our approximated universal fit (dotted contours) reproduces very well our full fit (continuous contours).
The best fit converged towards the SM; in particular data now disfavour the
solution with $c<0$ which appeared in previous fits.
Similar fits by the ATLAS and CMS collaborations are given in \cite{couplings}. The CMS result is similar to ours, while ATLAS has $c/a = 0.85^{+0.23}_{-0.13}$, due to their larger $h \to VV$ rates, which is compatible with our result at $1\sigma$ level.

The reason is visualised in the middle panel of fig.\fig{fitac}, where we show the bands favoured by the overall rates
for Higgs decay into heavy vectors ($WW$ and $ZZ$,
that get affected in the same way within the model assumptions), into fermions
($bb$ and $\tau\tau$, that get affected in the same way within the model assumptions)
and  into $\gamma\gamma$.
We see that these bands only cross around the SM point, $a=c=1$.
The full fit to all exclusive rates contains more information than this simplified fit.

In  the right panel of fig.\fig{fitac} we show the full $\chi^2$ restricted along the trajectories in the $(a,c)$ plane 
(plotted in the left panel)
predicted
by simple composite pseudo-Goldstone Higgs models 
in terms of the parameter $\xi= (V/F_\pi)^2$,
where $F_\pi$ is the scale of global symmetry breaking \cite{models}, $F_\pi \approx 130.4$~MeV.

\begin{figure}[t]
$$\includegraphics[width=0.45\textwidth]{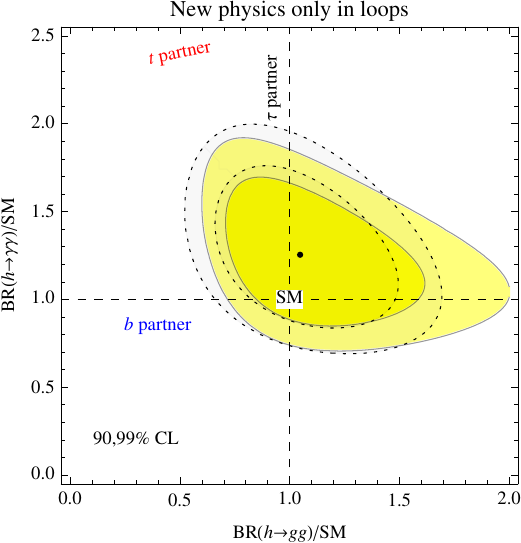}
\qquad
\includegraphics[width=0.45\textwidth]{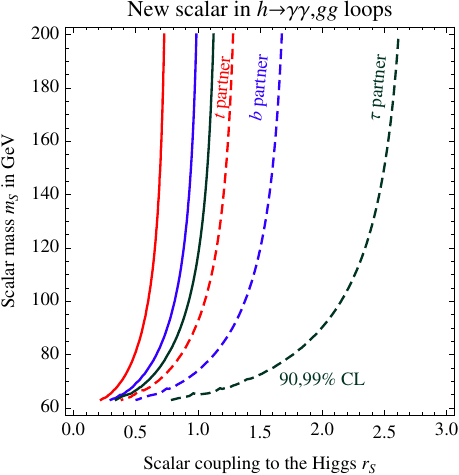}$$
\caption{\em {\bf Left}:  fit for the Higgs boson branching fraction to photons and gluons, with $1$ and $2 \sigma$ contours.
The dashed curves shows the possible effect of extra scalar partners of the top (red), of the bottom (blue), of the tau (black).
Dotted lines show the Gaussian approximation. 
{\bf Right}: Upper bound at $90\%$ (solid) and $99\%$ (dashed) C.L. on the new scalar coupling $r_S$ to the Higgs
 as a function of the new scalar mass $m_S$.
%\label{fig:fitBRfit-gsbound}
\label{fig:fitBR}}
\end{figure}

\begin{figure}[t]
$$\includegraphics[width=0.45\textwidth]{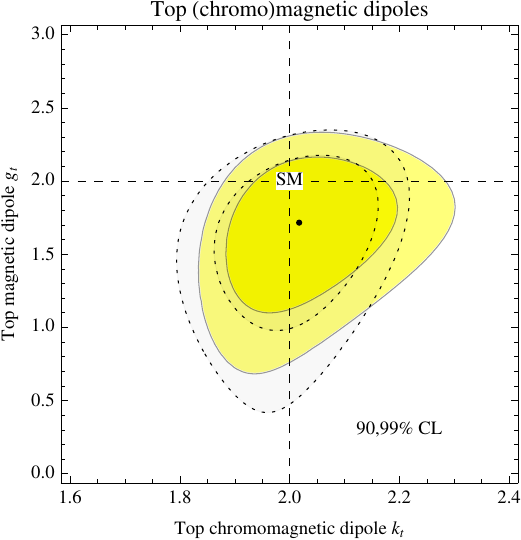} \qquad 
\includegraphics[width=0.46\textwidth]{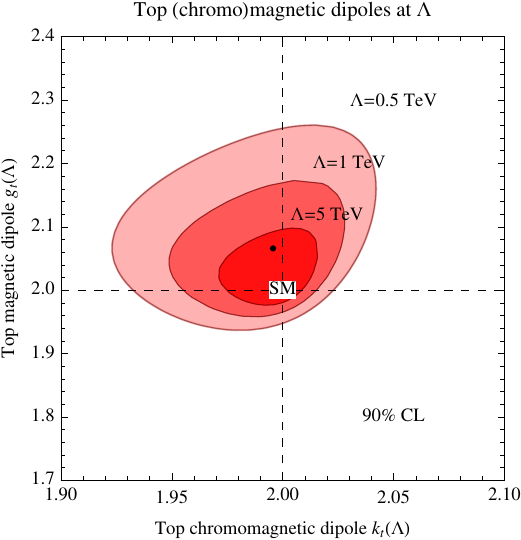}
$$
\caption{\em Best fit regions for the  magnetic and chromo-magnetic dipole moments of the top quark $g_t$ and $k_y$. Dotted lines show the Gaussian approximation.
{\bf Left}: as defined at $m_h$ according to the computation of~\cite{Labun:2012fg, Labun:2012ra}. {\bf Right}: as defined at a cutoff scale $\Lambda$ according to the computation of~\cite{Elias-Miro:2013gya}.
\label{fig:DipMom}}
\end{figure}

\begin{figure}[p]
$$\includegraphics[width=0.68\textwidth]{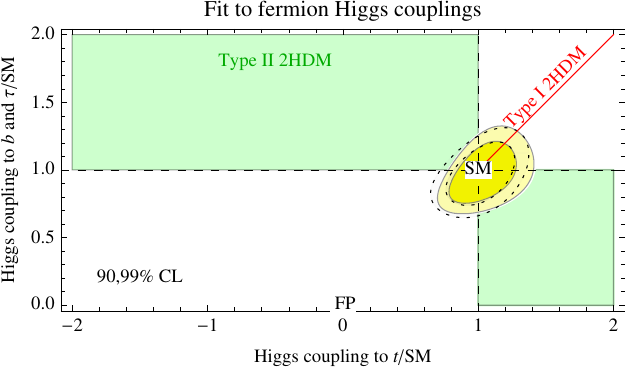}$$ 
$$\includegraphics[width=0.68\textwidth]{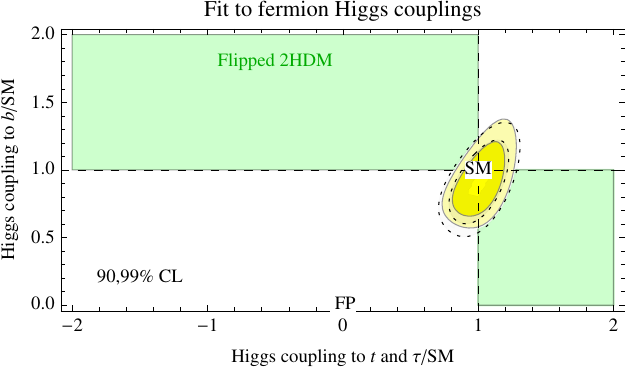}$$ 
$$\includegraphics[width=0.68\textwidth]{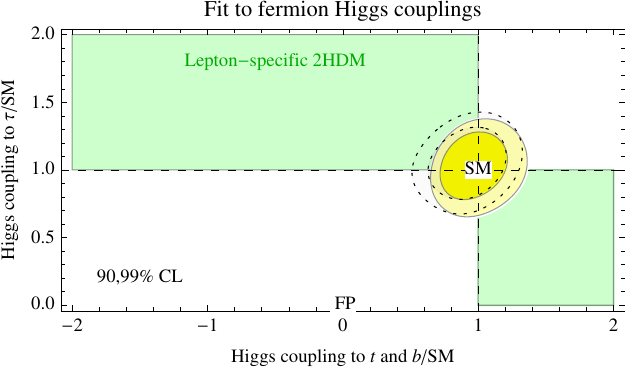}$$ 
\caption{\em 
Fit to the $t$-quark and to $b$-quark and $\tau$-lepton Yukawa couplings assuming 
the structure predicted by the various types of two Higgs doublet models. Dotted lines show the Gaussian approximation.
The point $(1,1)$ marked as `SM' is the Standard Model;
the point $(0,0$) marked as `FP' is the fermiophobic case.
}
\label{fig:fitbttau}
\end{figure}

\subsection{New physics only in the loop processes}\label{loops}
We assume here that only the loop processes are modified with respect to the SM predictions, summarized in appendix~\ref{app:L}.
This amounts to restricting our universal fit settings
\beq 
\label{BRfit}
r_t=r_b=r_\tau =r_\mu= r_W=r_Z=1,\qquad
\frac{ \Gamma(h\leftrightarrow gg)}{ \Gamma(h\leftrightarrow gg)_{\rm SM}}=r_g^2,\qquad
%\frac{\hbox{BR}(h\leftrightarrow gg)}{\hbox{BR}(h\leftrightarrow gg)_{\rm SM}}=r_g^2,\qquad
\frac{\Gamma(h\to\gamma\gamma)}{\Gamma(h\to\gamma\gamma)_{\rm SM}}=r_\gamma^2
%\frac{\hbox{BR}(h\to\gamma\gamma)}{\hbox{BR}(h\to\gamma\gamma)_{\rm SM}}=r_\gamma^2
\eeq
with BR$_{\rm inv}=0$ and $r_{Z\gamma}=1$. The latter assumption is at present justified because of the large experimental error 
in the  $h\to Z \gamma$ rate,
even though in general new physics in the loop processes would induce deviation from unity in both $r_{Z\gamma}$ and $r_\gamma$.  
The result is shown in the left panel of fig.\fig{fitBR}, in the form of a fit to the ratios of $\hbox{BR}(h \to gg)$ and $ \hbox{BR}(h\to \gamma \gamma)$ with respect to
the SM. One can see that the SM is well within the $1\sigma$ contour. The analogous ATLAS result~\cite{couplings} is instead
barely compatible with the SM at $2 \sigma$ level because they only fit ATLAS data, where $h \to VV$ rates have a central value above the SM.
The universal fit approximates the full fit reasonably well.
The dashed trajectories show the loop effect due to extra scalar particles with the same quantum numbers of the top (red), of the bottom (blue), of the tau (vertical black line).
The explicit expressions for the contribution of scalar, fermion and vector particles running in the loop can be found in appendix \ref{app:L}. 
Note that any additional colorless but electrically charged particle  would lead to the
same trajectory obtained for the scalar partner of the $\tau$.

\bigskip

To better investigate the constraints on a possible new scalar $S$, 
in the right panel of fig.\fig{fitBR} we show the upper bound,
as function of the scalar mass $m_S$, on the scalar coupling  $r_S$ to the Higgs boson,
defined by the coupling
\beq  r_S  \frac{2 m_S^2}{V}%g^{hSS}_{{\rm BSM}}   
h S S. \eeq
The resulting loop effects are summarised in  appendix~\ref{app:L}. The solid and dashed curves 
in fig.\fig{fitBR}b
are respectively  the upper bounds
at $90\%$ (solid) and $99\%$  (dashed) C.L. 
More stringent limits are obtained on the top and bottom partners than on the $\tau$ partner.

\bigskip

One can also use the universal fit with the assumption of eq.~(\ref{BRfit}) to derive indirect constraints 
on the top quark magnetic ($g_t$)
and chromomagnetic ($k_t$) dipole moments \cite{Labun:2012fg, Labun:2012ra}, which in the SM are expected to be 
respectively $g_t\approx 2$ and $k_t \approx 2$. 
Allowing $g_t$ and $k_t$ to vary freely,  the $h\rightarrow \gamma \gamma$ and $h\rightarrow g g$ amplitudes 
are modified with respect to the SM  as:
\begin{equation}\label{eq:chromo}
%\frac{A^{BSM}_{h\rightarrow \gamma \gamma}}{A^{SM}_{h\rightarrow \gamma \gamma}}
r_\gamma=\frac{ c_\gamma^{(W)}+c_\gamma^{(t)}  \left(  \frac{3}{8} g_t^2 -\frac{1}{2} \right)  } { c_\gamma^{(W)}+c_\gamma^{(t)}    }  ,
\qquad
r_g=\frac{3}{8} k_t^2 -\frac{1}{2} \,\,\, ,
\end{equation}
where the quantities $ c_\gamma^{(W)}$ and $c_\gamma^{(t)} $ are defined in eq.~(\ref{cloop:SM}) of the Appendix.
Numerically we have $ c_\gamma^{(W)}=-1.043$ and $c_\gamma^{(t)} =0.223$.
Fig. \ref{fig:DipMom} shows the $90\%$ and $99\%$ C.L. allowed regions for $g_t$ and $k_t$. 
The uncertainty on $k_t$ is comparable to the one from its direct measurements at the LHC and the Tevatron as combined 
in \cite{Kamenik:2011dk}, while the one for $g_t$ is even smaller. The conversion from the results of \cite{Kamenik:2011dk} 
is done in \cite{Labun:2012fg} for $g_t$, giving $-3.49 < g_t < 3.59$, and in \cite{Labun:2012ra} for $k_t$, giving $|k_t -2| < 0.2$ at $95\%$ C.L.

Eq.~(\ref{eq:chromo}) was computed by~\cite{Labun:2012fg, Labun:2012ra}
at the weak scale, in the phase with broken electroweak symmetry.
An analogous computation was performed in~\cite{Elias-Miro:2013gya}, 
promoting the dipoles to full  $\SU(2)_L\otimes{\rm U}(1)_Y$-invariant
effective operators with a non-renormalizable dimension $d>4$, suppressed by a factor $1/\Lambda^{d-4}$, $\Lambda$
being the cutoff of the theory. 
The result~\cite{Elias-Miro:2013gya} is that the dipole operators before electroweak symmetry breaking
contribute, via RGE mixing, to other one-loop suppressed operators affecting the $h\rightarrow \gamma \gamma$ and $h\rightarrow gg$ decay rates \cite{Manohar:2006gz}.  Finite parts are not computed.
Because of the RGE running from $\Lambda$ down to $m_h$, the effect is 
proportional to $\ln\Lambda/m_h$, differently from \eq{eq:chromo}.
Using the operator mixing result of \cite{Elias-Miro:2013gya} and parametrizing the $d=6$ dipole 
operators at $\Lambda$ via quantities analogous to 
$g_t$ and $k_t$ but defined at $\Lambda$, the decay rates \cite{Manohar:2006gz} can be written as
\beq
 r_\gamma=1 - \frac{  4/3  }{ c_\gamma^{(W)}+c_\gamma^{(t)}} \left(\frac{g_t(\Lambda)}{2} -1\right)  \log \frac{\Lambda}{m_h} \, , \,\, \,\,\,\,\,
r_g=1 - \frac{6 }{  c_{g}^{(t)} }   \left(\frac{k_t(\Lambda)}{2} -1\right)  \,  \log\frac{\Lambda}{m_h} \, ,
  \eeq
where the quantity $c_g^{(t)} $ is defined in eq.~(\ref{cloop:SM}) of the Appendix. Numerically $c_g^{(t)} =1.03$.
Repeating our fit, we obtain similar constraints as illustrated in the right panel of fig. \ref{fig:DipMom}, for 
representative values of the cutoff.

%

%%%%%%%%%%%%
\subsection{Models with two Higgs doublets}\label{sec:2HDM}
There are four types of two Higgs doublets models
(2HDM) where tree-level flavour-changing neutral currents (FCNCs) are forbidden by a $Z_2$ symmetry \cite{Glashow:1976nt} and both doublets $H_1$ and $H_2$
get a vacuum expectation value: 
\begin{itemize}
\item type I \cite{Haber:1978jt,Hall:1981bc} where only one doublet  couples to all quarks and leptons;
\item type II \cite{Hall:1981bc,Donoghue:1978cj},
where  up-type quarks couple to $H_{2}$ and $H_{1}$ couples to down-type quarks and leptons. The Higgs sector of the MSSM is a type II 2HDM;
\item type X (lepton-specific or leptophilic) where $H_{2}$ couples only to quarks and $H_{1}$ couples only to  leptons;
\item type Y (flipped) \cite{2hdmtXY}, where $H_{2}$ couples to up-type quarks and $H_{2}$ to down-type quarks, and (contrary to the type II HDM) leptons couple to $H_{2}$.

\end{itemize}
For an extensive review see \cite{Branco:2011iw} and for some previous fits see \cite{2HDMfits}.
The modification to Yukawa couplings to up-type and down-type quarks and leptons in the four 2HDMs
are: 
\begin{center}
\begin{tabular}{lcccc}
  & Type I & Type II & Type X (lepton-specific) & Type Y (flipped) 
  \\[3mm]
  $r_t$ 
    & $\displaystyle \phantom{-}\sfrac{\cos \alpha}{\sin \beta}$ 
    & $\displaystyle \phantom{-}\sfrac{\cos \alpha}{\sin \beta}$ 
    & $\displaystyle \phantom{-}\sfrac{\cos \alpha}{\sin \beta}$
    & $\displaystyle \phantom{-}\sfrac{\cos \alpha}{\sin \beta}$
  \\	
  $r_b$  
    & $\displaystyle \phantom{-}\sfrac{\cos \alpha}{\sin \beta}$ 
    & $\displaystyle -\sfrac{\sin \alpha}{\cos \beta}$ 
    & $\displaystyle \phantom{-}\sfrac{\cos \alpha}{\sin \beta}$
    & $\displaystyle -\sfrac{\sin \alpha}{\cos \beta}$
  \\
  $r_\tau$  
    & $\displaystyle \phantom{-}\sfrac{\cos \alpha}{\sin \beta}$ 
    & $\displaystyle -\sfrac{\sin \alpha}{\cos \beta}$ 
    & $\displaystyle -\sfrac{\sin \alpha}{\cos \beta}$
    & $\displaystyle \phantom{-}\sfrac{\cos \alpha}{\sin \beta}$
\end{tabular}
\end{center}
As usual, $\tan \beta = v_{2}/v_{1}$ is the ratio of the VEVs of the two doublets and $\alpha$ is the mixing angle of the CP-even mass eigenstates.  The SM limit corresponds to $\beta-\alpha = \pi/2$. In all of the models the vector couplings are also modified as
\begin{equation}\label{sab}
r_W = r_Z = \sin (\beta - \alpha).
\end{equation}
The results of our fits are presented in fig.\fig{fitbttau} in terms of the  fermion couplings
$r_t,r_b,r_\tau$,
restricted by the 2HDM models to lie within the green regions. (We do not show the region for $r_{b,\tau} \approx 1$ which is allowed since the measurements have no sensitivity to the signs of these couplings.) We find that in each case, it is $r_t$ that dominates the fit and the bottom contributions to gluon fusion and $h \to \gamma\gamma$ are negligible.The effect of the charged Higgs boson in the $h \to \gamma \gamma$ loop is neglected.

The type II 2HDM (upper panel) allows for independent modification of the $t$ coupling $r_t$, and for a common modification of the $b$ and $\tau$ couplings, $r_b=r_\tau$.  The former is predicted be reduced and the latter enhanced by the model. 
The modification of \eq{sab} of the vector couplings can be equivalently written
as $r_W = r_Z = (1+r_tr_b)/(r_t+r_b) \simeq 1 + \epsilon_t\epsilon_b/2$, showing that it is
a small second order effect.
In this model a negative $t$ Yukawa coupling is still allowed at slightly more than 99\%~CL.
The red line in the same panel shows the parameter space allowed by
type I 2HDM, where all the couplings scale uniformly.

In the flipped 2HDM (middle panel) the $\tau$ Yukawa coupling changes in the same way as the $t$ coupling and the region with negative coupling is disfavoured by data. 
Finally, in the leptophilic 2HDM (lower panel) the $t$ and $b$ couplings vary in the same way, while the $\tau$ coupling is independent.

The universal fit provides a reasonable approximation to the full fit in all 2HD models.

\subsection{Supersymmetry}\label{MSSM}
\begin{figure}[t]
$$\includegraphics[width=0.45\textwidth]{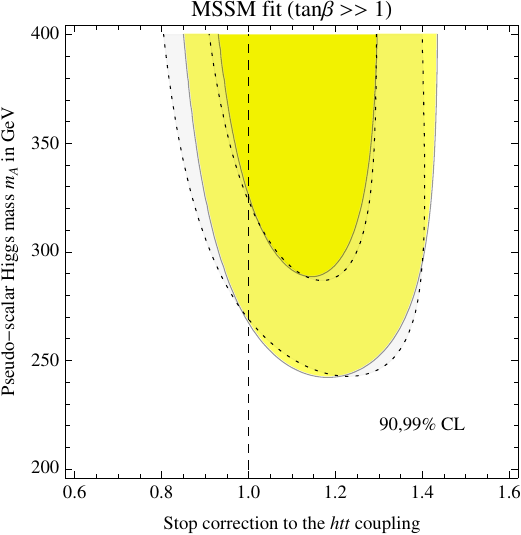}
\qquad\includegraphics[width=0.45\textwidth]{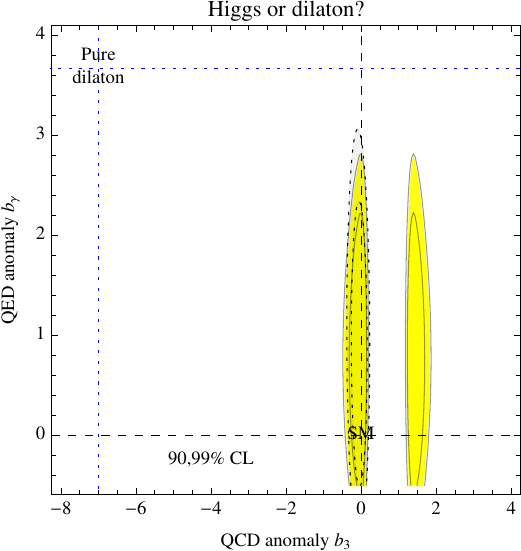}$$ 
\caption{\em {\bf Left}: Fit to the two main effects present in supersymmetry:
stop loop correction to the $ht\bar{t}$ coupling and tree-level modification of
the Higgs couplings due to the two-Higgs doublet structure. Dotted lines show the Gaussian approximation.
{\bf Right}:  fit as function of the $\beta$-function coefficients $b_3=b_\gamma$ that parameterise dilaton models.
The SM Higgs is reproduced at the experimentally favored  point $b_3=b_\gamma=0$, while
the pure dilaton is excluded at more than $5\sigma$.
\label{fig:sigma}
\label{fig:fitbt}}
\end{figure}

Supersymmetry can affect Higgs physics in many different ways, such that it is difficult to make general statements.
We here focus on the two most plausible effects:
\begin{itemize}
\item  The stop squark loop  affect the $h\leftrightarrow gg,\gamma\gamma,Z\gamma$ rates.
Given that the stop has the same gauge quantum numbers of the top, such effects are correlated
and equivalent to a modification of the Higgs coupling to the top (as long as it is not directly measured via
the $t\bar t h$ production cross section)
by an amount given by
\beq 
R_{\tilde{t}} = 1+\frac{m_t^2}{4}\left[\frac{1}{m_{\tilde{t}_1}^2}+
\frac{1}{m_{\tilde{t}_2}^2}-\frac{(A_t-\mu/\tan\beta)^2}{m_{\tilde{t}_1}^2m_{\tilde{t}_2}^2}\right]
\eeq
in the limit of heavy stop masses, $m_{\tilde{t}_{1,2}}\gg m_t$.
Notice that $R_{\tilde{t}}$ can be enhanced or reduced with respect to one, depending on the latter mixing term.

\item The type II 2HDM structure of  supersymmetric models
modifies at  tree-level   the Higgs couplings, as already discussed in section~\ref{sec:2HDM}.
\end{itemize}
All of this amounts to specialise the universal $\chi^2$ inserting
the following values of its parameters
\beq r_t = R_{\tilde{t}}\frac{\cos\alpha}{\sin\beta},\qquad
r_b=r_\tau=r_\mu =-\frac{\sin\alpha}{\cos\beta},\qquad
r_W = r_Z = \sin(\beta-\alpha).\eeq
Furthermore, the parameters $r_g,r_\gamma, r_{Z\gamma}$ relative to loop processes are fixed as in \eq{eq:loops}.
We trade the $\alpha$ parameter (mass mixing between Higgses)  for the pseudo-scalar Higgs mass $m_A$ using
\beq \tan2\alpha = \frac{m_A^2 + M_Z^2}{m_A^2 - M_Z^2}\tan2\beta.  \eeq
Finally, we assume a large $\tan\beta$, as motivated by the observed value of the Higgs mass (a large $\tan\beta$ amplifies the stop contribution to the Higgs mass).
The left panel of fig.\fig{sigma} shows the resulting fit.   Once again, the universal fit is an adequate approximation of the full fit.
Of course, supersymmetry can manifest in extra ways not considered here, e.g.\ very light staus or charginos could enhance $h\to\gamma\gamma$~\cite{Carena}.
 
\subsection{Data prefer the Higgs to the dilaton}\label{sec:dilaton}
As another example of a model where both the tree-level and the loop level
Higgs couplings are modified, we consider the dilaton.
The dilaton is an hypothetical particle $\varphi$, that, like the Higgs, couples 
to  SM particles with strength proportional to their masses~\cite{radion}.
More precisely the dilaton has a coupling
 to the trace of the energy-momentum tensor 
$T_{\mu\nu}$, suppressed by some unknown scale $\Lambda$:
\beq \frac{\varphi}{\Lambda} T_\mu^\mu =  \frac{\varphi}{\Lambda} \left(\sum_f m_f \bar f f - M_Z^2 Z_\mu^2 - 2M_W^2 W_\mu^2 
+b_3\frac{\alpha_3}{8\pi} G_{\mu\nu}^aG_{\mu\nu}^a +b_\gamma \frac{\alpha_{\rm em}}{8\pi} F_{\mu\nu}F_{\mu\nu}
\right)
\label{radion} . \eeq
The dilaton couplings to  $gg$ and $\gamma\gamma$ differ from the corresponding Higgs boson couplings,
because \eq{radion} contains the latter two quantum terms, that are present in $T_{\mu}^\mu$ because
scale invariance is anomalous and broken at quantum level by the running of the couplings.
Indeed  $b_3$ and $b_\gamma$ are the $\beta$-function coefficients
of the strong and electromagnetic gauge couplings. 
In the SM they have the explicit values
$b_3=-7$ and $b_\gamma=11/3$: we call  `pure dilaton' this special model,
which gives  a significant enhancement of $h\leftrightarrow gg$.

Models where a dilaton arises usually often contain also
new light particles, such that $b_3$ and $b_\gamma$ can differ from their SM values.
Thereby we perform a generic fit where $b_3$ and $b_\gamma$
are free parameters in addition to $\Lambda$.
Then, our universal fit is adapted to the case of the generic dilaton by setting 
\bea
r\equiv r_W = r_Z = r_t = r_b = r_\tau = \frac{V}{\Lambda},\qquad
r_g \approx r (1-1.45 b_3),\qquad r_\gamma\approx r(1+0.15 b_\gamma)
\label{radionrate}
\eea 
where $V=246\GeV$.

In our previous analyses~\cite{Giardino:2012ww,Giardino:2012dp}, the dilaton 
gave fits of comparable quality to the SM Higgs,  despite the significantly different predictions of the dilaton:
enhanced $\gamma\gamma$ rates and reduced vector boson fusion rates.
The first feature is no longer favoured by data, and the second feature is now disfavoured: so
we find that  present data  prefer the Higgs to the `pure dilaton' at about $7\sigma$ level.
We then consider the generic dilaton, showing in
fig.\fig{sigma}b that the allowed part of its parameters space is the one
where it  mimics the Higgs,
possibly up to a sign difference in $r_g$ and/or $r_\gamma$. The linear couplings of the dilaton in \eq{radion}
become identical to those of the SM Higgs in the limit $b_3=b_\gamma=0$ and $\Lambda = V$.
This situation is not easily realisable in models, given that adding extra charged particles
increases $b_\gamma$ rather than reducing it; one needs to subtract particles by e.g.\ assuming that
that 3rd generation particles are composite~\cite{brando}.

The universal approximation works reasonably well, although it cannot reproduce these disjoint solutions.

\begin{figure}[t]
$$\includegraphics[width=0.45\textwidth]{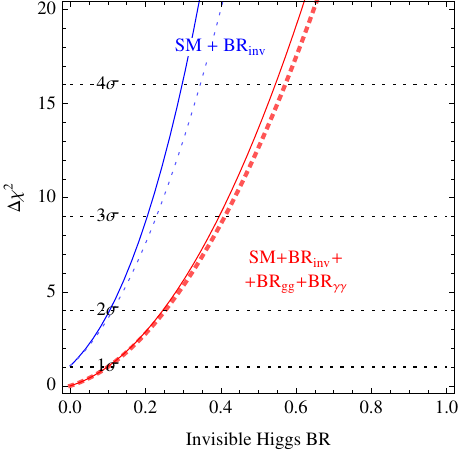}\qquad
\includegraphics[width=0.47\textwidth]{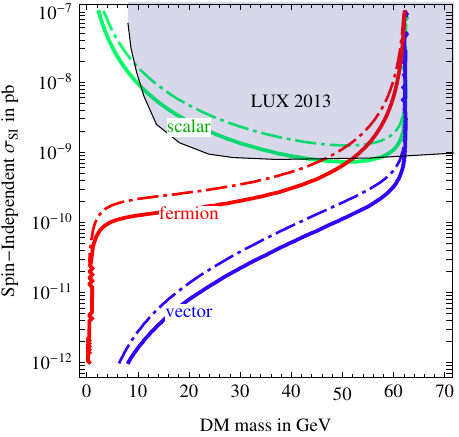} $$
\caption{\em  {\bf Left:} fits to the invisible Higgs boson branching fraction under the two different assumptions
described in section~\ref{inv} for DM which directly couples to the Higgs.
The full fit (continuos curves) is in reasonable agreement with the universal fit (dotted curves). 
{\bf Right}:
upper limit on the spin-independent DM cross section on nucleons as a function of the DM mass for scalar (green), Majorana fermion (red) and vector (blue) DM directly coupling to the Higgs.
We adopted the 95\% C.L. bounds  ${\rm BR}_{\rm inv}<0.22$ (solid, \eq{BR1}) and $<0.34$ (dot-dashed, \eq{BR2}). 
The shaded region is excluded at $90\%$ C.L.\  by  {\sc LUX}2013 \cite{Aprile:2012nq}.
\label{fig-sXmX}
\label{fig:fitinv}}
\end{figure}

\subsection{Higgs boson  invisible width}\label{inv}

Next, we allow for a Higgs boson  invisible width, for example into Dark Matter (this does not comprise undetectable decays into known physics, such as Higgs to light jets).%
\footnote{
Note that such decays are only undetectable at hadron colliders due to large QCD backgrounds and trigger problems, but could be detected at an $e^{+}e^{-}$ collider in the $ZH$ production mode.
}
We perform two fits.
\begin{enumerate}
\item In the first fit, the invisible Higgs width is the only new physics.
We find (blue curves in fig.\fig{fitinv}a) that present data imply 
$\hbox{BR}_{\rm inv} = -0.12\pm0.12$.
The one-sided upper bound, computed restricting to $0\le \hbox{BR}_{\rm inv}\le 1$, is
\beq \label{BR1}
\hbox{BR}_{\rm inv} < 0.17\hbox{ at 95\% C.L.}\eeq

\item In addition to the invisible width we also  allow for non-standard values of
$h\to \gamma\gamma$ and $h\leftrightarrow gg$, finding a
weaker constraint on BR$_{\rm inv}$ (red curves in  fig.\fig{fitinv}a)
\beq\label{BR2}
\hbox{BR}_{\rm inv} < 0.26\hbox{ at 95\% C.L.}  \eeq
The reason is that an enhanced $gg\to h$ production rate can partially compensate for an invisible Higgs width,
but a full compensation would be possible only by enhancing all production rates by the same amount.
The Higgs coupling to vectors is independently measured to agree with SM predictions from electroweak precision data.

\end{enumerate}
Notice that the main constraint for BR$_{\rm inv}$
does not come from the direct search for
$pp\to Z h\to \ell\ell \slashed{E}_T$ (included in our data-set)
but from the global fit~\cite{Giardino:2012ww,inv}.

%%%%%%%%%%%%%%%
\subsection{Dark Matter models}\label{sec:DM}
The  invisible Higgs boson decay width \cite{inv} constrains
Dark Matter (DM) candidates with mass below $M_h/2$. 
The Higgs sector of the SM allows for a direct coupling to particles of a hidden sector.
%The determination of the properties of the Higgs boson allows to constrain hidden sector particles. 
If the latter are stable and interact weakly with the SM sector, they could represent viable Dark Matter (DM) candidates. 
If DM particles have mass below $M_h/2$, the Higgs boson can thus decay into a pair of DM particles, which would escape detection.
Invisible Higgs decays are constrained by the fact that the ATLAS and CMS Higgs rates are compatible with the predictions of the SM Higgs boson.
The experimental bound on BR$_{\rm inv}$ can be used  to constrain the DM mass and its elastic cross section on nucleons
probed in direct detection experiments, 
as illustrated for instance in  \cite{Djouadi:2012zc},
where DM is assumed to be either a scalar $S$, or a Majorana fermion $f$ or a vector $V$ coupled to the Higgs as
\begin{eqnarray} 
 r_S  \frac{2 m_S^2}{V}%g^{hSS}_{{\rm BSM}}   
h S S 
+ r_f \frac{ m_f}{V}  %g^{h\bar f f}_{{\rm BSM}}   
h \bar f f 
+ r_V  \frac{2 m_V^2}{V} %g^{hVV}_{{\rm BSM}}   
h V_\mu V_\mu   
%+ \cdots 
\,\, .
\label{eq:efflag}
\end{eqnarray}
The partial Higgs decay width into dark matter $\Gamma(h \rightarrow {\rm DM}\, {\rm DM})$ and the spin-independent ${\rm DM}$-proton elastic cross section
$\sigma_{\rm SI}$ can be  calculated in terms of the parameters of the above Lagrangian.  
Both are proportional to the square of the DM-Higgs coupling, so that the 
ratio $\mu\equiv  \sigma_{\rm SI}/\Gamma(h \to {\rm DM}\, {\rm DM})$
depends only on the the unknown DM mass and on the known masses and couplings of the relevant SM particles
(see for instance the expressions provided in~\cite{Djouadi:2012zc}). 

This allows us to relate the invisible Higgs branching fraction to the DM direct detection cross section:
\begin{equation}
\hbox{BR}_{\rm inv}\equiv \frac{\Gamma(h \to {\rm DM}\,{\rm DM})}{\Gamma^{\rm SM}_h+\Gamma(h \to {\rm DM}\,{\rm DM})} = \frac{\sigma_{\rm SI}}{ \mu{\Gamma^{\rm SM}_h}+\sigma_{\rm SI}}
\end{equation}
where $\Gamma^{\rm SM}_h=4.1\MeV$ is the total Higgs decay width into all SM particles, 
that we fix to its SM prediction.
For a given DM mass, an upper bound on the Higgs invisible branching fraction implies an upper bound on the DM scattering cross section
on nucleons. The relation between the invisible branching fraction and the direct detection cross section strongly depends on the
spinorial nature of the DM particle, in particular, the strongest (weakest) bound is derived in the vectorial (scalar) case.

Imposing  the  upper bounds on BR$_{\rm inv}$ derived in section~\ref{inv},
fig.~\ref{fig-sXmX} shows the corresponding
upper limits on the spin-independent DM cross section on nucleons as a function of the DM mass, 
in the case of scalar (green), Majorana fermion (red) and vector (blue) DM candidates.

%This figures is inspired to the fig. in ref  \cite{Djouadi:2012zc}.
%It generalizes the results already obtained in sec 4.4 of ref \cite{Giardino:2012ww}. 
%[If I understood correctly one does not even need to assume thermal relic abundance... I will check this.]

In all cases, the derived bounds are stronger than the direct one from LUX2013 as long as the mass of ${\rm DM}$ is lighter than $M_h/2$.
This conclusion does not rely on the assumption that DM is a thermal relic that reproduces 
the observed cosmological DM abundance.
The limit on $\sigma_{\rm SI}$ crucially depends on the assumption that DM directly couples to the Higgs.
Larger values of $\sigma_{\rm SI}$ remain possible in different models, where DM couples to the $Z$
or directly to nucleons via loops of supersymmetric or other particles.

%and would only be stronger if ${\rm DM}$ constitutes only a fraction of DM in the universe.  CHE SIGNIFICA???

%Of course, if the mass of ${\rm DM}$ is heavier than $M_h/2$, the Higgs boson cannot decay into DM, in which case the LHC cannot
%compete with the XENON bounds.

%%%%%%%%%%%%%%%%%%%%%%%%%%%%%%%%%%%%%%%

\section{Discussion and Conclusions}
\label{sec:concl}

The LHC experiments reported 
their measurements of  Higgs boson properties at the Moriond 2013 conferences, based on the full collected luminosity during 2011 and 2012. At the same time, Tevatron reported their final Higgs results. 
With the crucial inclusion of the full CMS $\gamma\gamma$ data (missing in previous analyses),
at this stage all main Higgs results from Tevatron and from the first phase of LHC have  been basically presented.
Those results will drive our understanding of particle physics, until new 13~TeV LHC data will be available. 

\smallskip

Motivated by these results,  we have performed a state-of-the-art global fit to Higgs boson data, including all sub-categories
studied by the experimental collaborations, for a total of 56 experimental inputs, as summarised in fig.\fig{data}. 
We found that  the average Higgs rate is $0.99\pm 0.09$ in SM units,  supporting  the SM Higgs boson hypothesis.
The Higgs boson mass is usually determined from the peaks in the invariant mass distribution
of $ZZ$ and $\gamma\gamma$.  We  performed the first measurement of the Higgs boson mass from the rates, 
finding that the two determinations are compatible:
\beq M_h = 
\left\{\begin{array}{ll}
125.15\pm0.24~\GeV& \hbox{from the peaks,}\\
125.0\pm 1.8~\GeV  & \hbox{from the rates.}\\
\end{array}\right. 
\eeq
The LHC physics program has been  successful: with only $\approx 25$/fb of data per experiment 
the  Higgs boson has been discovered and several of its properties determined  within 
$\approx \pm 20\%$ precision.
We are now entering into the era of precision Higgs physics --- deviations from 
the SM due to new physics no longer can dominate the data.
This observation allowed us to propose a `universal' form in which experiments could report their results 
allowing
theorists to easily test any desired model.
The new assumption that makes possible this significant simplification is that new physics is a small correction to the SM.
While we used all publicly available data to present our own global combination in `universal' form in \eq{universal},
we stress that only the experimental collaborations can perform a fully precise analysis,
for example including the correlations among experimental uncertainties.

\smallskip

%We applied this methodology to perform a series of model independent as well as model dependent fits.
We studied several new physics scenarios beyond the SM. 
We determined from data the production cross sections (assuming standard Higgs decays) and
the Higgs decays widths (assuming standard productions), finding that they lie along the SM predictions.
In a more general context, we allowed all possible Higgs boson couplings to any SM particle to
deviate from its SM value, finding that couplings to the
$W,Z,t,b,\tau$ must lie around their SM predictions up to uncertainties of about $\pm20\%$
(see fig.~\ref{fig:fitcouplings}b).
In particular, non-standard Higgs boson couplings to vectors,
predicted by composite Higgs models, are most stringently constrained.
The scenario of negative Higgs coupling to fermions (`dysfermiophilia') that gave the best fit with early LHC data is now  disfavoured at
more than $2\sigma$.

%Similarly, the best fits to $h\to\gamma\gamma$ and $h\to gg$ that also showed significant deviations from the SM values
%in earlier fits are now comfortably compatible with the SM.

We considered various specific new physics models:  new scalars, 2HDM, supersymmetry, dilaton, composite Higgs,
invisible Higgs decays, possibly into Dark Matter particles, anomalous couplings of the top, etc.
The results of those fits are presented in numerous figures throughout the paper. 
Qualitatively,  all reach the same conclusions: 
\begin{itemize}
\item[i)] best fit regions lie along SM predictions, imposing constraints on new physics;
\item[ii)] our simple ÕuniversalÕ fit is a reasonable approximation to the full fit.
\end{itemize}
In particular we find that, with the latest data, the dilaton alternative to the Higgs is now excluded at $5\sigma$, 
with the exception of the special non-minimal dilaton tuned to exactly reproduce the Higgs (section~\ref{sec:dilaton}).

%Unfortunately this experimental success does not provide us any hint of new physics beyond the SM. To the contrary, the LHC has confirmed the low scale SM.
%In the light of that, perhaps the most important implications of the LHC concern the high energy model building at scales $\Lambda \sim {\cal O}(10^{10})$~GeV
%where the Higgs quartic coupling vanishes and induces vacuum instability.  

\bigskip

% We will update this paper at the light of future results.
We will update this paper when future results become available.
\medskip

\paragraph{Acknowledgement} 
This work was supported by the ESF grants  8943, MTT8, MTT59, MTT60, MJD140 by the recurrent financing SF0690030s09 project
and by  the European Union through the European Regional Development Fund. The work of P. P. G. has been partially funded by the ``Fondazione A. della Riccia''.
This work was supported in part by the European Programme	PITN-GA-2009-23792 (UNILHC).

%%%%%%%%%%%%%%%%%%%%%
\appendix\small

\section{New physics contributions to loop processes}

%%%%%%%%%%%%%%%%%%%%%%%%%%%%%

\label{app:L}
%The coefficients $g_p^{\rm SM}$ of the first line of eq. (\ref{eq:efflag}) represent the Higgs coupling to the particle $p$ in the SM, 
%so that deviations from unity of their coefficients $R_p$ represent effects of new physics beyond the SM. 
The  coefficients in the second line of \eq{Lh} arise at one-loop. They are obtained by summing the contributions of all scalars ($S$)
fermions ($f$) and vectors ($V$) that couple to the Higgs as in \eq{eq:efflag}.
%Deviations from unity of the coefficients $r$ 
%of these lines would be due to effects of new physics beyond the SM. 
%The extra terms in eq.~(\ref{eq:efflag}) accounts for possible beyond-SM scalars, fermions and vectors that couple to the Higgs boson. 
%It is useful to define the beyond-SM couplings to the Higgs of the third line of eq. (\ref{eq:efflag}) in terms of SM quantities,
%$g^{hpp}_{{\rm BSM}} = r_p g^p_{\rm SM}$, where for a particle $p=S,f,V$ the couplings $g_{\rm SM}^p$ arise at tree-level 
%and are defined as 
%\begin{equation}
% g^S_{\rm SM} = \frac{2 m_S^2}{v}\,\,, \,\, g^f_{\rm SM} = \frac{ m_f}{v}  \,\,, \,\, g^V_{\rm SM} = \frac{2 m_V^2}{v} \,\,, 
%\end{equation}
%for a scalar, fermion or vector, respectively. 
The explicit expressions for the loop effects are~\cite{rev}:
\begin{eqnarray}
c_g^{(S)} =  \frac{C_2^S }{2}  r_S A_S(\tau_S) \qquad c_g^{(f)} =  2 C_2^f r_f A_f(\tau_f) \,\,\,\,\,\,\,\,\,\, \,\,\,\,\,\,\,\,\, \,\,\,\,\,\,\,\,\,\,\,\,\,\,\,\,\\ 
c_\gamma^{(S)} =  \frac{N_S Q_S^2}{24} r_S A_S(\tau_S) \qquad 
c_\gamma^{(f)} =  \frac{N_f Q_f^2}{6} r_f A_f(\tau_f) \qquad 
c_\gamma^{(V)} =  -\frac{7 Q_V^2}{8} r_V A_V(\tau_V) \,\, \nonumber
\end{eqnarray}
where for each particle $p=S,f,V$, $\tau_p=m_h^2/ 4 m_p^2$,
$N_p$ is the number of colors, $C_2^p$ is the Casimir of the color representation (Tr$(T^a T^b)=C_2 \delta^{ab}$),
and the loop functions are
\begin{eqnarray}
A_S(\tau) &=& \frac{3}{\tau^2} \left[ f(\tau) -\tau\right] \,\, , \,\, 
A_f(\tau) = \frac{3}{2 \tau^2} \left[ (\tau -1) f(\tau) + \tau \right] \\
A_V(\tau)& =& \frac{1}{7 \tau^2} \left[3 (2\tau -1) f(\tau) + 3 \tau + 2 \tau^2 \right] 
\end{eqnarray}
with $f(\tau) = \arcsin^2(\sqrt{\tau}) \,\,\, {\rm for} \,\,\, \tau \le 1$ such that
$A_p(\tau_p) \to 1$ in the limit $\tau_p \to 0$ (heavy $p$-particle).

In particular, in the SM, the $hgg$ coupling is dominated by the top loop, and the $h\gamma\gamma$ coupling arise from the sum of the top and $W$ boson loops: 
\begin{equation}
%g^{hgg}_{\rm SM} = \frac{\alpha_s}{12 \pi v} c_g^{\rm SM}  \,\,, \,\,  
c^{gg}_{\rm SM} = c_g^{(t)} = A_f(\tau_t) \qquad
%g^{h\gamma \gamma}_{\rm SM} = \frac{\alpha}{ \pi v} c_\gamma^{\rm SM}  \,\,, \,\, 
c^{\gamma \gamma}_{\rm SM}= c_\gamma^{(t)} +c_\gamma^{(W)} =  \frac{2}{9} A_f(\tau_t) -\frac{7}{8} A_V(\tau_W) \,\, . 
\label{cloop:SM}
\end{equation}
Beyond the SM (BSM) physics affects the parameters $r_g$ and $r_\gamma$ as
\begin{equation}
r_g =  1+  \frac{ c^{gg}_{{\rm BSM}}   }{   c^{gg}_{\rm SM} }  ,\qquad
r_\gamma = 
 1+  \frac{ c^{\gamma\gamma}_{{\rm BSM}}   }{   c^{\gamma\gamma}_{\rm SM} } .
\label{eq:rgrg}
\end{equation}
For example,  additional scalar particles with the same quantum numbers of a stop, sbottom and stau respectively
 contribute to $c^{gg}_{{\rm BSM}}$   
and to $c^{\gamma\gamma}_{{\rm BSM}} $  as:
\beq\begin{array}{lll}\displaystyle
c_g^{(\tilde t)} = \frac{1}{4} r_{\tilde t} A_S(\tau_{\tilde t}) & \displaystyle
c_g^{(\tilde b)} = \frac{1}{4} r_{\tilde b} A_S(\tau_{\tilde b}) &\displaystyle
c_g^{(\tilde \tau)} =0 \\[3mm]   
c_\gamma^{(\tilde t)} = \frac{1 }{18} r_{\tilde t} A_S(\tau_{\tilde t})  & \displaystyle
 c_\gamma^{(\tilde b)} = \frac{1 }{72} r_{\tilde b} A_S(\tau_{\tilde b}) & \displaystyle
c_\gamma^{(\tilde \tau)} =\frac{1 }{24}  r_{\tilde \tau} A_S(\tau_{\tilde \tau}) .
% \frac{N(r_S) Q_S^2}{24} R_S  A_S(\tau_S)
\end{array}\eeq

\bigskip\bigskip

%
%\begin{equation}
%R_g^{(\tilde t)} = \left|  1+  \frac{  R_{\tilde t} A_0(\tau_{\tilde t})   }{ A_{1/2}(\tau_t) }  \right|\,\,\,\, , \,\,\, 
%R_\gamma^{(\tilde t)} = \left|  1+  \frac{ R_{\tilde t}  A_0(\tau_{\tilde t})}{  A_{1/2}(\tau_t) +\frac{3}{4} A_{1}(\tau_W) }  \right| 
%\end{equation}
%\begin{equation}
%R_g^{(\tilde b)}
%= \left|  1+  \frac{  R_{\tilde b} A_0(\tau_{\tilde b})    }{ A_{1/2}(\tau_t) }  \right| \,\,\,\, , \,\,\, 
%R_\gamma^{(\tilde b)} 
%= \left|  1+  \frac{ \frac{ 1 }{4} R_{\tilde b}  A_0(\tau_{\tilde b})}{   A_{1/2}(\tau_t) +\frac{3}{4} A_1(\tau_W) }  \right| 
%\end{equation}
%\begin{equation}
%R_g^{(\tilde \tau)} = 1  \,\,\, , \,\,\,
%R_\gamma^{(\tilde \tau)} 
%= \left|  1+  \frac{ \frac{ 3}{4} R_{\tilde \tau}  A_0(\tau_{\tilde \tau})}{  A_{1/2}(\tau_t) +\frac{3}{4} A_1(\tau_W) }  \right| 
%\end{equation}

%%%%%%%%%%%%%%%%%%%

\footnotesize

\begin{multicols}{2}

\end{multicols}
\end{document}